\documentclass[twocolumn]{openjournal}
\usepackage{amsmath}
\usepackage{bm}
\usepackage{url}
\usepackage{graphicx}
\usepackage{hyperref}

\makeatletter
\let\UrlSpecialsOld\UrlSpecials
\def\UrlSpecials{\UrlSpecialsOld\do\/{\Url@slash}\do\_{\Url@underscore}}%
\def\Url@slash{\@ifnextchar/{\kern-.02em\mathchar47\kern-.1em}%
   {\kern-.02em\mathchar47\kern-.02em\penalty\UrlBigBreakPenalty}}
\def\Url@underscore{\nfss@text{\leavevmode \kern.06em\vbox{\hrule\@width.3em}}}
\makeatother

\renewcommand{\vec}[1]{\mbox{\boldmath$#1$}}

\newcommand{\mat}[1]{\mbox{\boldmath{$\mathsf #1$}}}
\newcommand{\vhat}[1]{\vec{\hat{#1}}}

\newcommand{\origami}{{\scshape origami}}

\newcommand{\R}{\mathcal{R}}
\newcommand{\avg}[1]{\left\langle{#1}\right\rangle}

\newcommand{\hgpc}{\,$h^{-1}$\,Gpc}

\newcommand{\LCDM}{$\Lambda$CDM}

\newcommand{\eg}{e.g., }

\chardef\til=`\~

\shorttitle{Halo Spin from Primordial Inner Motions}
\shortauthors{Mark Neyrinck et al.}

\begin{document}

\title{Halo Spin from Primordial Inner Motions}

%\author[Mark C.\ Neyrinck et al.] {
%{\parbox{\textwidth}{
\author{Mark C.\ Neyrinck}
\affil{Ikerbasque, the Basque Foundation for Science}
\affiliation{Dept. of Theoretical Physics, University of the Basque Country, Bilbao, Spain}
\author{Miguel A.\ Aragon-Calvo}
\affiliation{Instituto de Astronom\'{i}a, UNAM, Apdo.\ Postal 106, Ensenada 22800, B.C., M\'{e}xico}
\author{Bridget Falck}
\author{Alexander S.\ Szalay}
\affiliation{Department of Physics and Astronomy, The Johns Hopkins University, Baltimore, MD 21218, USA}
\author{Jie Wang}
\affiliation{National Astronomy Observatories, Chinese Academy of Science, Datun Road, Beijing, PR China}

%\small
%$^{1}$Ikerbasque, Basque Foundation for Science\\
%$^{2}$Dept. of Theoretical Physics, University of the Basque Country, Bilbao, Spain\\
%$^{3}$Instituto de Astronom\'{i}a, UNAM, Apdo.\ Postal 106, Ensenada 22800, B.C., M\'{e}xico\\
%$^{4}$Department of Physics and Astronomy, The Johns Hopkins University, Baltimore, MD 21218, USA\\
%$^{5}$National Astronomy Observatories, Chinese Academy of Science, Datun Road, Beijing, PR China
%}
%$^{3}$DIPC, UPV/EHU, E-48080 San Sebasti\'{a}n, Spain}

%\author{Mark C. Neyrinck\\
%{\rm \small 
%Department of Physics and Astronomy, The Johns Hopkins University, Baltimore, MD 21218, USA}
%}

%\date{\today}

\begin{abstract}
The standard explanation for galaxy spin starts with the tidal-torque theory (TTT), in which an ellipsoidal dark-matter protohalo, which comes to host the galaxy, is torqued up by the tidal gravitational field around it. We discuss a complementary picture, using the relatively familiar velocity field, instead of the tidal field, whose intuitive connection to the surrounding, possibly faraway matter arrangement is more obscure. In this `spin from primordial inner motions' (SPIM) concept, implicit in TTT derivations but not previously emphasized, the angular momentum from the gravity-sourced velocity field inside a protohalo largely cancels out, but has some excess from the aspherical outskirts. At first, the net spin scales according to linear theory, a sort of comoving conservation of familiar angular momentum. Then, at collapse, it is conserved in physical coordinates. Small haloes are then typically subject to secondary exchanges of angular momentum.

The TTT is useful for analytic estimates. But a literal interpretation of the TTT is inaccurate in detail, without some implicit concepts about smoothing of the density and tidal fields. This could lead to misconceptions, for those first learning about how galaxies come to spin. Protohaloes are not perfectly ellipsoidal and do not uniformly torque up, as in a naive interpretation of the TTT; their inner velocity fields retain substantial dispersion. Furthermore, quantitatively, given initial conditions and protohalo boundaries, SPIM is more direct and accurate than the TTT to predict halo spins.  We also discuss how SPIM applies to rotating filaments, and the relation between halo mass and spin, in which the total spin of a halo can be thought of as a sum of random contributions.
\end{abstract}

\maketitle

\section{Introduction}
The topic of how galaxies come to be spinning, and what determines that spin, is fundamental. This paper explores basic explanations for this process. We start with the standard pedagogical explanation, the tidal torque theory (TTT), explained briefly in the next paragraphs. But we also describe a complementary explanation, which we find to have some advantages.

\citet[][P69]{Peebles1969} introduced the TTT explanation for galaxy spin into the standard cosmological picture of structures sourced from primordial density fluctuations. It goes as follows: as a protohalo collapses to form a halo, it does so embedded in a gravitational tidal field, produced by the external arrangement of matter. Generally, a protohalo is aspherical, with a moment of inertia not perfectly aligned with tidal field, and thus the tidal field produces a torque on it. (For this paper, we use `spin' synonomously with `angular momentum.')  We discuss this process for collisionless dark-matter haloes, neglecting processes that could cause galaxy spins to differ from that of their host haloes.

\citet[][W84]{White1984} made a further important contribution, stating that the TTT had already `come to be accepted.'  Here we briefly review W84's arguments. The full expression of a protohalo's spin in physical coordinates (not comoving with the expansion of the universe) is \citep{Doroshkevich1970}
\begin{equation}
\vec{L}(a) = \rho_0 a^5 \int_V (\vec{x}(\vec{q})-{\vec{\bar{x}}}) \times {\vec{v}} d^3 q.
\label{eqn:fullang}
\end{equation}
Here, $\vec{x}$ is the comoving Eulerian coordinate at some time, a function of $\vec{q}$, the Lagrangian coordinate, i.e.\ the position in the initial conditions. $\rho_0$ is the mean density, $a$ is the scale factor, tracking the expansion of the Universe, and $V$ is the volume of the protohalo in the initial conditions.

In the TTT argument, particle velocities are treated as ballistic, retaining their direction, but with speeds increasing according to the \citet[][ZA]{Zeldovich1970} approximation, linear theory in Lagrangian coordinates. In this approximation, the velocity $\vec{v}=-\vec{\nabla}{\phi(\vec{q})}$, the (Lagrangian-coordinates) gradient of the gravitational potential at a particle's initial position $\vec{q}$.  Neglecting displacements from the initial conditions (i.e.\ replacing $\vec{x}$ with $\vec{q}$), 
\begin{equation}
\vec{L}(a) = -a^2\dot{D}\bar{\rho}_0 \int_V (\vec{q}-{\vec{\bar{q}}}) \times \vec{\nabla}{\phi(\vec{q})} d^3 q.
\label{eqn:zeldl}
\end{equation}
Here $D(a)$ is the growth rate, which tracks the linear-theory amplitude of fluctuations, and equals $a$ in a matter-dominated universe. \citet[][P02]{PorcianiEtal2002} approximate $a^2 \dot{D}\approx D^{3/2}$, which holds well in flat cosmologies. In a matter-dominated epoch, this is a factor of $a^{3/2}$, as used by W84. One factor of $a$ can be thought to come from the scaling from comoving to physical coordinates. Another factor of $a^{1/2}$ gives the physical-coordinates velocity scaling in the linear regime. This $a^{1/2}$ factor is well-known to users of {\scshape Gadget} \citep{Gadget2} particle velocities; it makes them constant in code units, in the linear regime.

Expanding the potential to second order in $\vec{q}$ (see W84) gives the TTT expression,
\begin{equation}
L_i(a) = a^2 \dot{D}(a)\epsilon_{ijk}T_{j\ell}I_{\ell k},
\label{eqn:ttt}
\end{equation}
where $L_i$ is the $i$th component of $\vec{L}$, $\mat{T}$ is the Hessian of the potential at the protohalo center, $T_{ij}=\partial\Phi/\partial x_i\partial x_j$, and $\mat{I}$ is the inertia tensor of the protohalo.

The TTT works rather well as an estimate of a halo's spin \citep{CatelanTheuns1996,SugermanEtal2000}, but there are subtleties, e.g.\ care must be taken to smooth the tidal field at the appropriate scale (P02). It is also not necessarily clear how to define the protohalo center. (The volume centroid?  The potential minimum? The maximum of the density, smoothed on some scale?)  Also, misconceptions are possible if the TTT is taken too literally. The idea that an ellipsoid uniformly spins up in a non-rotating background of the same density is inaccurate at the interface, with collisions in some locations, and evacuation in others. Protohaloes are not delineated by any obvious boundary in the initial conditions, ellipsoidal or otherwise; they can be potato-shaped and peakless \citep{LudlowPorciani2011}, sometimes needing a full $N$-body simulation for an accurate estimate of boundaries.

P69 introduced the TTT to the now-standard picture of galaxies arising from initial small perturbations. The TTT was originally proposed in the steady-state theory \citep{Hoyle1949,Sciama1955}, in which a literal interpretation, with distant galaxies exerting torques, may have been more applicable. But, in our opinion, the idea of misaligned tidal tensors (obscurely related to possibly distant matter) and moment-of-inertia tensors acting on a perfect ellipsoid is rather abstract and hard to visualize for an introductory textbook-level explanation of the fundamental question of how galaxies spin. Many learners would realize that this explanation is an approximation, not to be taken too literally, but in our opinion, the following complementary, more accurate picture could be useful.

We call this picture `Spin from Primordial Inner Motions' (SPIM), following Eq.\ (\ref{eqn:zeldl}). The motions are referred to as `primordial inner\footnote{`Inner' may be misleading, since with irrotational initial velocities, the net spin comes from the aspherical shell just inside the boundary of the protohalo, as explained in \S \ref{sec:kelvin}. But in general, the full velocity field inside the protohalo contributes.}' because we emphasize the role of the primordial velocity field inside the protohalo, gravitationally sourced by the density field inside the protohalo, and around it. To our knowledge, SPIM has not previously been advocated as a conceptual framework for halo spin, even if it accords with some experts' intuition.

The SPIM prescription is as follows: sum up the angular momentum from the velocity field inside a protohalo in the initial conditions. Extrapolate it according to the growth factor (written simply as $a$ here for simplicity) for halo spin until collapse, $L(a)\propto a^{3/2}$. This $L(a)\propto a^{3/2}$ describes a sort of comoving conservation of angular momentum, which (as we show below) holds well in the median. After collapse, treat $L(a)$ as conserved in physical coordinates. This is pedagogically appealing since it involves only the initial velocity field, more readily conceived of than the tidal field smoothed on the scale of the protohalo, and its relation to the inertia tensor. The initial velocity field is a first-order derivative (the gradient) of the potential, but the tidal field is second-order (the Hessian). But it should be kept in mind that this is a complementary, not competing, viewpoint, describing essentially the same picture as TTT. SPIM inner motions ultimately come from gravity, just like tidal fields. These are both relevant to shear in the velocity field \citep{JonesVdw2009}, which is often relevant to spin generation, as well.

SPIM could be useful for predicting final spins in approximate cosmological realizations, if the patches that collapse are known, using e.g. the adhesion model \citep[e.g.][]{GurbatovEtal2012,HiddingEtal2012,NeyrinckEtal2018,HiddingEtal2019,Hidding2019}. Another use of SPIM (not calling it that) is in the {\scshape pinocchio} algorithm \citep{MonacoEtal2002} for predicting halo catalogs from initial conditions uses Lagrangian positions for parts of its halo spin estimate. A quantitative test of a prescription similar to SPIM was by P02, who tested the TTT against the ZA. They found that the ZA was better at predicting halo spins than the TTT, not surprising since the TTT is an approximation to the ZA. But SPIM is simpler than the ZA; it does not require explicit particle displacements, but only the initial velocity field in the protohalo.

Secondary processes (e.g.\ orbital angular momentum in mergers) often contribute substantially to spin; indeed, accretion of spin via satellites has even been proposed as a source of halo spin \citep{VitvitskaEtal2002}. This sort of picture is appropriate if one wants to track a collapsed object as it grows and merges with smaller objects in time. But if what is tracked is the protohalo, i.e.\ all the matter that ends up in the final halo, for very large haloes (with $M>10^{13-14}M_\odot$), primary processes dominate. For smaller haloes, secondary processes may be substantial.  An additional, crucial complication is that a galaxy's spin typically differs from its host dark-matter halo spin \citep[e.g.\ ][]{LiaoEtal2017,GaneshaiahEtal2019}, possibly because of dissipative processes in the gas and star-formation feedback \citep{EfstathiouJones1980}.

Correlations of the velocity field are simpler than correlations of the tidal field, so this SPIM picture may be useful to study spin-spin correlations and intrinsic galaxy alignments \citep[\eg][]{JoachimiEtal2015}. We also hope it will be useful to investigate how the cosmic web influences galaxy halo spin \citep[\eg][]{AragonCalvoEtal2007,CodisEtal2012,LibeskindEtal2012,StewartEtal2013,TrowlandEtal2013,AragonCalvoYang2014,GaneshaiahEtal2018,CodisEtal2018,KraljicEtal2019}.

This SPIM idea might seem disconcertingly reminiscent of the vortical primeval cosmic turbulence theory \citep{Wiseguy1951,Gamow1952,Ozernoi1972}, a once-popular theory to explain galaxy angular momentum that is now disfavored. It was this theory that P69 argued against when first proposing the TTT. The gravity-sourced motions that we emphasize could be called `irregular motions (``initial'' angular momentum),' as P69 put them, in stating that they would be neglected. But he meant the vortical gas eddies of the primeval turbulence theory, not motions sourced by gravity. See also \citet{Peebles2018} for a recent perspective on this history.

Still, it is subtle how halo rotation arises from an irrotational (vorticity-free) flow. Usually, the primordial velocity field is thought to be irrotational, because perturbatively, expansion dampens vorticity. At later epochs, vorticity generally arises in patches with multiple collisionless dark-matter streams \citep{PichonBernardeau1999, HahnEtal2015}; \citet{WangEtal2014} discuss vorticity generation from a different perspective. But of course, the accumulation of spin in a finite patch is different from vorticity generation, as we discuss further in Section \ref{sec:kelvin}.

The structure of this paper is as follows. In \S \ref{sec:2D}, we start with an investigation in 2D, convenient for building intuition because the spin is simply a scalar. It also is relevant to investigating spin of 3D filaments. In \S \ref{sec:particlespin}, we look at how individual particles contribute to halo spins. In \S \ref{sec:halospin}, we look at the behavior of total halo spin with time. In particular, in \S \ref{sec:halospinzeld}, we show that halo spin predictions in SPIM are nearly the same as those in the more conceptually and computationally complex ZA. In \S \ref{sec:randomwalk}, we show how considering particle contributions individually helps to understand the scaling of spin with mass in the initial conditions. In \S \ref{sec:kelvin}, before concluding, we explain how rotation can arise in an aspherical object from an irrotational flow; all of the angular momentum comes from the object's outskirts.

\section{Two dimensions}
\label{sec:2D}
We start our analysis of halo spin with a 2D $N$-body dark-matter (i.e.\ gravity only) simulation, further simplifying the physical problem by suppressing small-scale fluctuations in the initial conditions. In the next couple of paragraphs, we give details of the simulation and halo finding, and then analyze our results.

The 2D $N$-body simulation has 1024$^2$ particles, and a box size of 32 Mpc/$h$. The initial conditions are a 2D slice of particles from a $1024^3$ grid generated via the ZA with a BBKS \citep{BBKS} initial power spectrum, with modes of wavelength $< 1$ Mpc/$h$ suppressed in amplitude, to further simplify the problem. The ($x$-$y$) plane of particles was replicated along the $z$-axis to match the number of slices in the original particle grid. However, we reduced the mass resolution of the replicated planes with increasing distance from the (central) plane in a similar way as done in zoom-in simulations in order to reduce the computational cost.  It was run using a version of {\scshape Gadget2} \citep{Gadget2}, modified to calculate forces and update positions only in the $x$ and $y$ directions. Because of this replication along the $z$-axis, each particle represents a cylinder, interacting with other cylinders effectively with a 2D version of gravity. For the initial power spectrum and expansion history, we used a generic \LCDM\ ($\Omega_{\rm M}=0.3$, $\Omega_\Lambda=0.7$, $\sigma_8=0.8$, $h=0.7$) set of cosmological parameters.

\begin{figure}
  \begin{center}
     (Lagrangian) \hspace{0.3cm} Particle's contrib.\ to halo spin \hspace{0.3cm} (Eulerian)\\
    \includegraphics[width=\columnwidth]{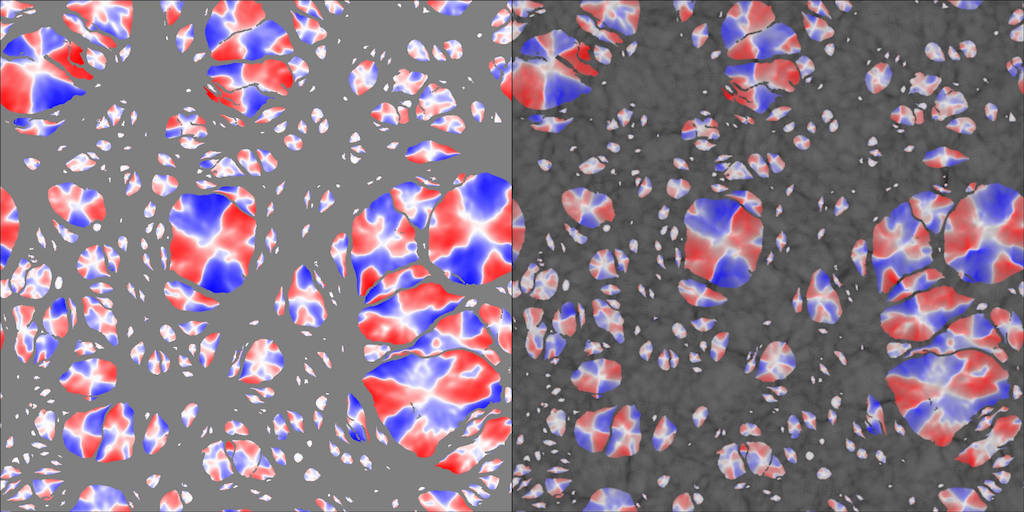}
    \includegraphics[width=\columnwidth]{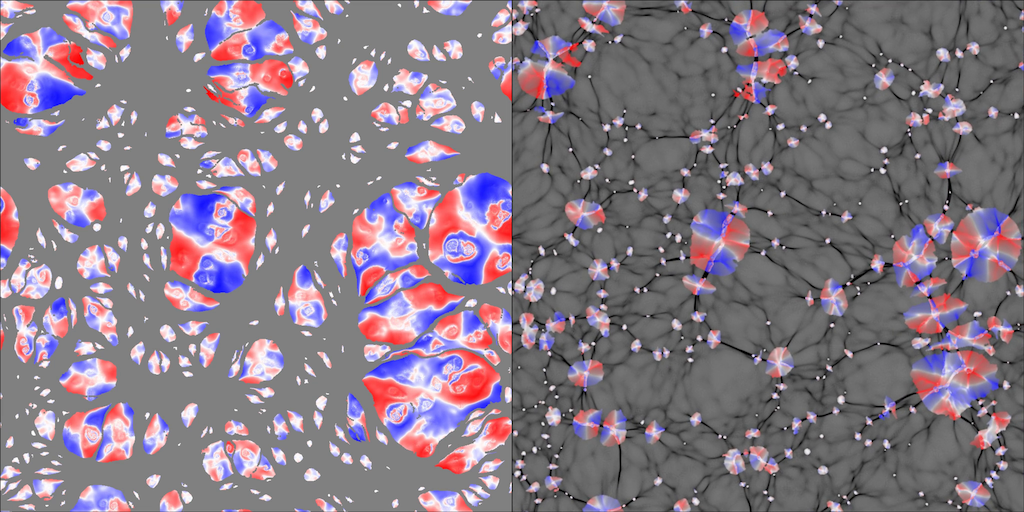}
    \includegraphics[width=\columnwidth]{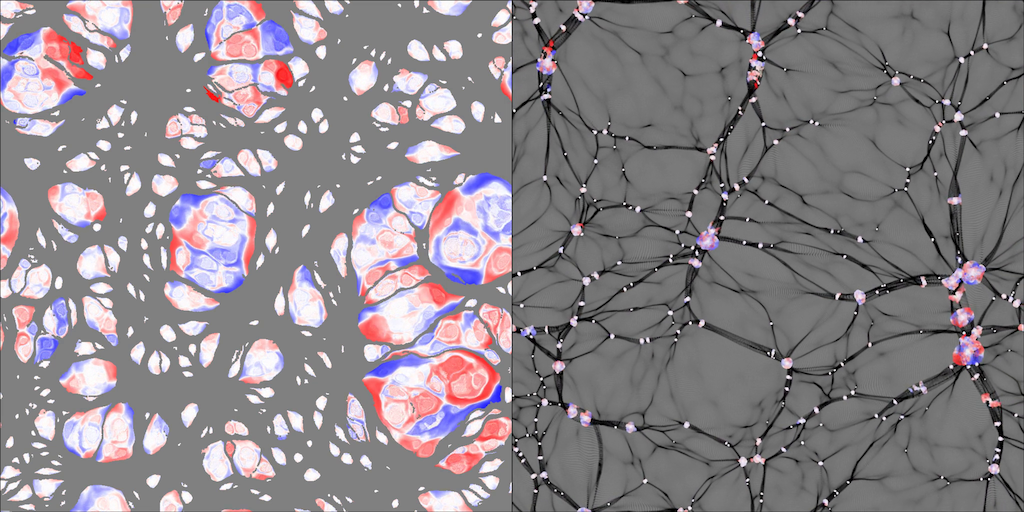}
  \end{center}  
  \caption{In initial (left) and final (right) coordinates of a 2D $N$-body simulation, the contributions of each halo particle to the angular momentum of its halo, at three snapshots of time. Red, counterclockwise particles contribute positively; blue, clockwise particles contribute negatively. We use a nonlinear colorscale to enhance small deviations from zero (white). In greyscale, blue appears darker than red. Black particles are not in haloes above the mass cut, 20 particles. Notice that for particles in halo outskirts in the middle panels, the colors are similar as in the top panels. Time advances from top to bottom; the snapshots are at $a=0.09$, 0.3, and 1. Animation at \url{https://www.youtube.com/watch?v=7KjesL_hP7c&list=PLJxZSP5_6rBmo7Rz8JgnE1eI2bxRGXGCT}.}
  \label{fig:locangmom}
\end{figure}

\begin{figure}
  \begin{center}
   (Lagrangian space) \hspace{1cm} Halo spin \hspace{1cm} (Eulerian space)\\
    \includegraphics[width=\columnwidth]{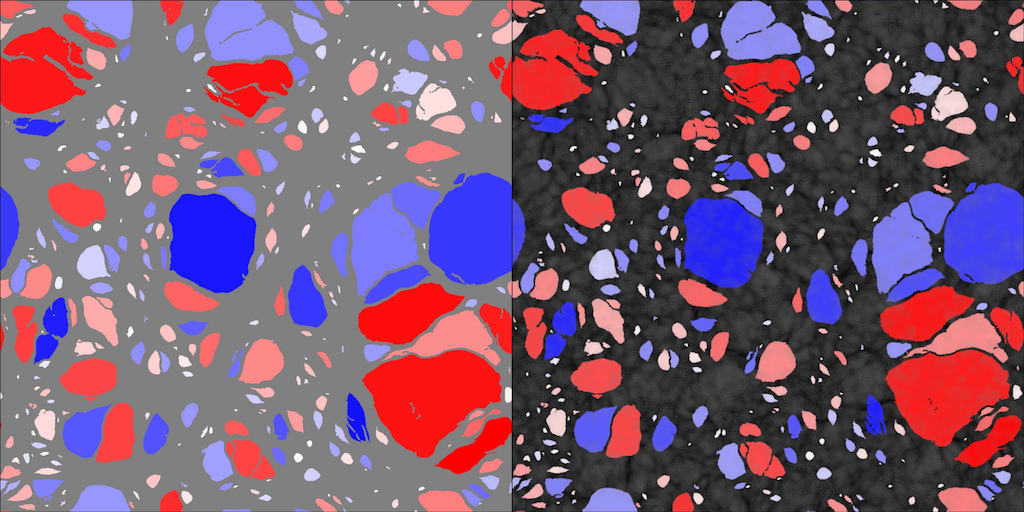}
    \includegraphics[width=\columnwidth]{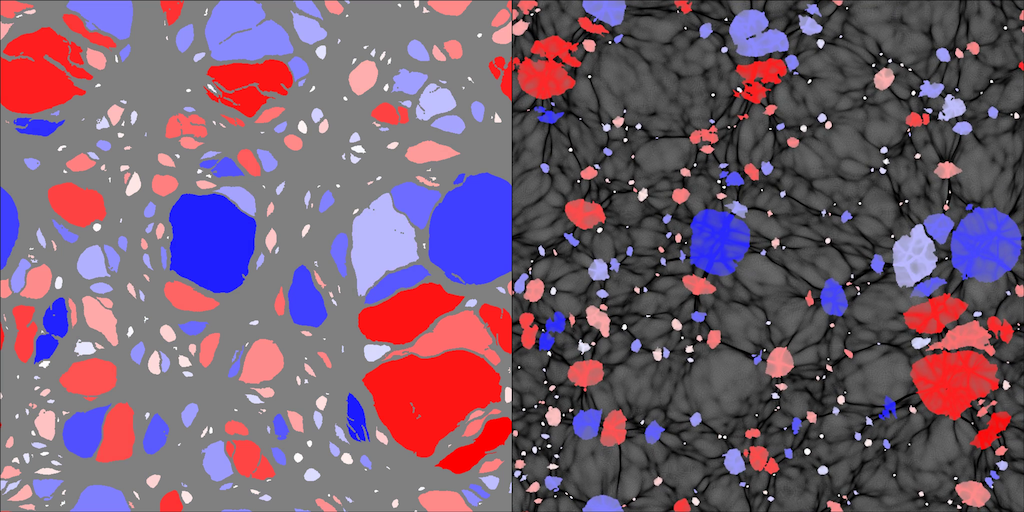}
    \includegraphics[width=\columnwidth]{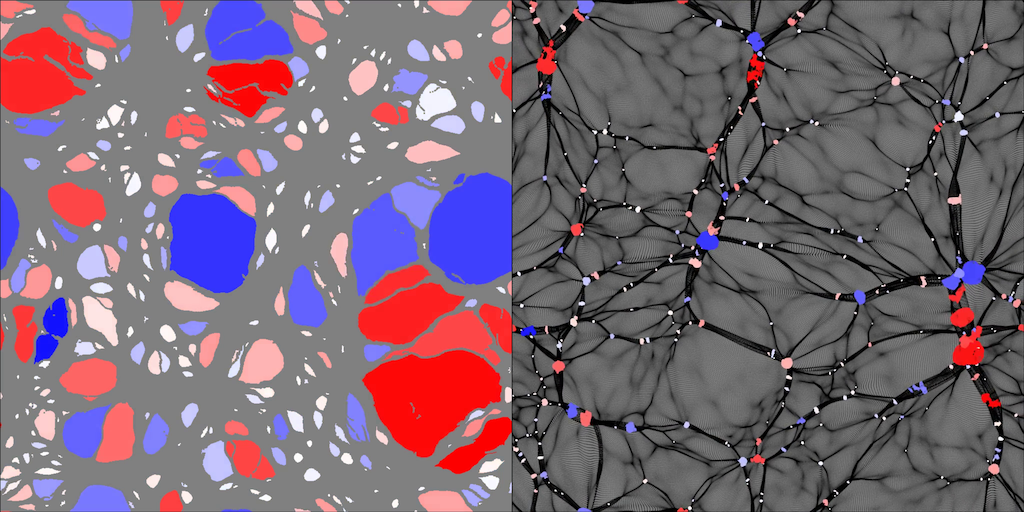}
  \end{center}  
  \caption{As in Fig.\ \ref{fig:locangmom}, coloring by the total angular momentum of a particle's halo. Animation at \url{https://www.youtube.com/watch?v=dmqJX_rSnKE&list=PLJxZSP5_6rBmo7Rz8JgnE1eI2bxRGXGCT}.}
  \label{fig:spins_lageul}
\end{figure}

We detect haloes in the 2D simulation using the \origami\ algorithm \citep{FalckEtal2012}. In \origami, a particle is classified as a halo particle if, going from the initial to final conditions, it has `folded' (crossed some other particle) along two (in 2D) initial orthogonal axes. We then join together groups of particles adjacent on the initial Lagrangian square grid to form haloes\footnote{Several haloes returned in this process were actually groups of haloes apparently distinct by eye, joined by small bridges of halo particles. The spurious fragmentation in $N$-body simulations with truncated initial power \citep{WangWhite2007} likely contributed to this. Applying a mathematical morphology erosion operator cut these bridges, returning haloes that in almost all cases corresponded to visual expectation. Erosion, used e.g.\ by \citet{PlatenEtal2007}, shaves off cells within a specified distance (here, 1 grid spacing) of a boundary from all contiguous blobs. We computed the erosion by smoothing the Lagrangian `halo' (1) and `not-halo' (0) field with a circular top-hat filter of radius 1 pixel; after smoothing, we classified as `halo' particles all corresponding pixels with a value $>0.99$.}.

How does each particle contribute to its halo's spin? Fig.\ \ref{fig:locangmom} shows the contribution $L_i/m_{\rm part}=(\vec{x_i}-\vec{\bar{x}})\times(\vec{v_i}-\vec{\bar{v}})$ of each halo particle $i$ to the spin of its halo, where $m$ is the particle mass. To get a physical spin, the quantity shown should be scaled by $a^{3/2}$. To clearly see both extreme and low-amplitude fluctuations, we use a $\sinh^{-1}$-scaled variable, $L_{\rm colorscale}=L_{\sinh}\sinh^{-1}(L_i/L_{\sinh})$, where $L_{\sinh}/m=10^{3} {\rm km/s~Mpc}/h$. This responds linearly for $|L_i| \ll L_{\sinh}$, and logarithmically for $|L_i| \gg L_{\sinh}$.

There is a characteristic quadrupolar alternating pattern in most of the haloes, previously discussed analytically and illustrated by  \citet{CodisEtal2015}. In the figure, and especially in the accompanying animation, one might see that many of the quadrupolar spin patterns are mainly driven by collapse along the first axis of the halo. Flows toward the first axis of collapse show up as contributing clockwise or counter-clockwise spin; one of these wins out, somewhat randomly. This pattern is also similar to vorticity patterns in filament cross sections \citep{LaigleEtal2013} in Eulerian space at low redshift, after smoothing. Both quadrupolar patterns represent similar flows, but this spin field is nonzero even in the initial, vorticity-free initial conditions. Also, note that, as discussed in \S \ref{sec:kelvin}, the irrotational flow should guarantee that the spin contributed from inside a maximal circle fitting inside a halo should vanish. In particular, the area centroid of each halo lies on a white contour, where $L=0$, in Fig.\ \ref{fig:locangmom}.

The `haloes' in this 2D simulation can be thought of essentially as cross-sections of filaments in a 3D simulation. This implies that we should generically expect that filaments have some net rotation along the axis, as suggested by \citet{LaigleEtal2013}. Filament rotation also generically happens in an origami, or tetrahedral-collapse approximation \citep{Neyrinck2016iau,Neyrinck2016tet}, in which filaments are extrusions of 2D `origami twist folds;' that was a motivation for this 2D study.

What happens when we sum up these particle-by-particle contributions? Fig.\ \ref{fig:spins_lageul} shows the same, but coloring each halo particle by its halo's total angular momentum. Most haloes retain their color with time.

\begin{figure}
  \begin{center}
    \includegraphics[width=\columnwidth]{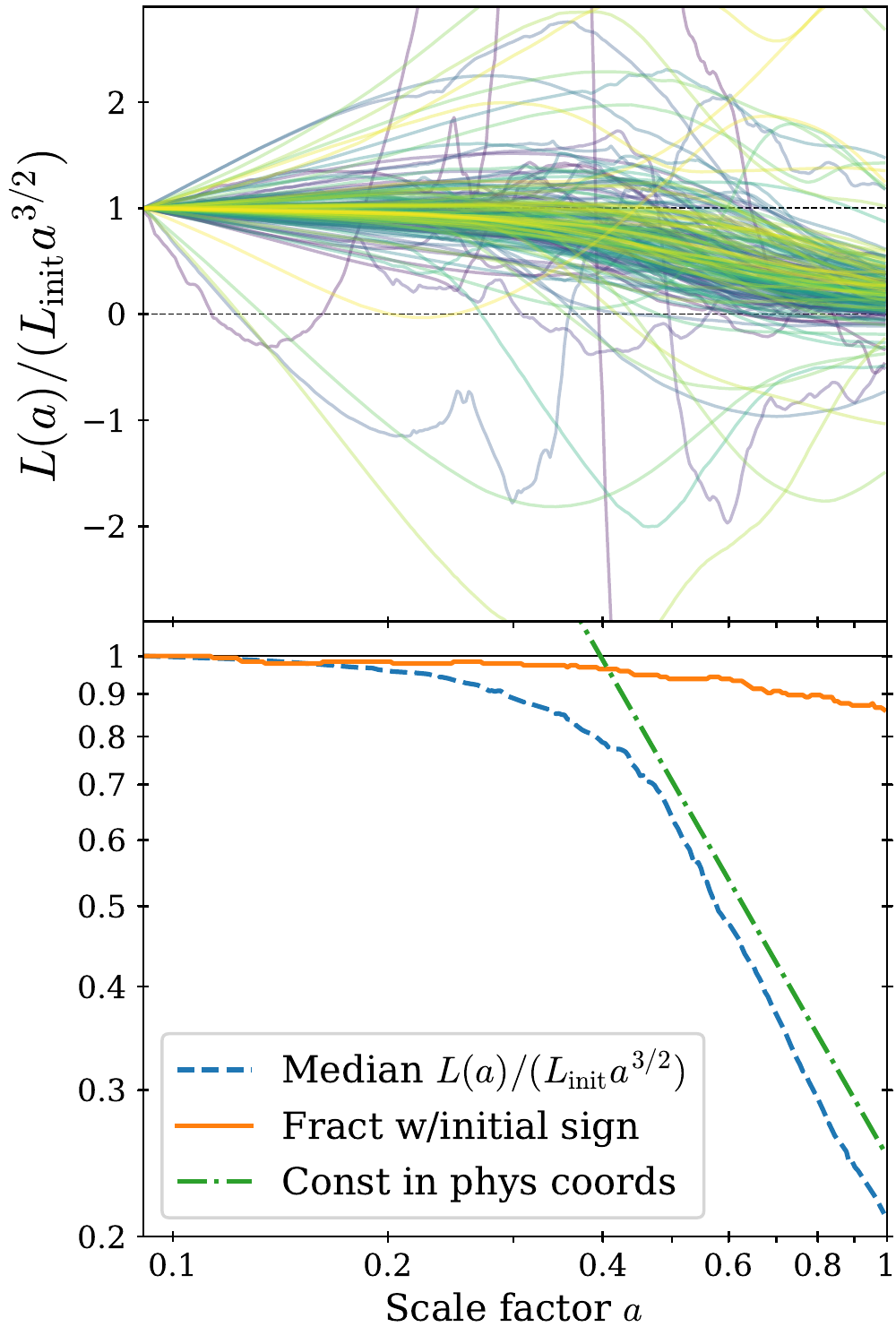}
    \end{center}  
  \caption{\textbf{\textit{Top}}: Trajectories in time of halo angular momentum $L(a)$, normalized by their initial values $L_{\rm init}$, in a 2D simulation. Purple (dark), green, and yellow (light) trajectories show low, middle, and high-mass haloes. \textbf{\textit{Bottom}}: The median behavior, after dividing by $a^{3/2}$, is constant at early times, transitioning at $a\sim 0.3$ to a constant in physical coordinates, i.e. following a line $\propto a^{-3/2}$. 14\% of halo spins switch sign from the initial to final snapshot.}
  \label{fig:hawobblo_2d}
\end{figure}

Fig. \ref{fig:hawobblo_2d} shows trajectories in angular momentum, $L(a)$, for haloes of at least 100 particles. There is a substantial spread in behavior, but in the median, the particles behave as in linear theory, with $L(a)\propto a^{3/2}$ up to $a\sim0.3$. Then, spin is conserved in physical coordinates, giving constant $L(a)$ ($\propto a^{-3/2}$, in this plot). This general behavior was already found by W84, and it will appear below in the 3D results, as well.

\section{Particle contributions to 3D spin}
\label{sec:particlespin}
In 3D, the behavior of halo spins is similar as above. We analyze a simulation, 201, from the Indra suite of simulations (Falck et al., in prep), a set of 512 independent simulations with 1024$^3$ particles and box size 1\hgpc. It uses a WMAP7 \LCDM\ set of cosmological parameters ($\Omega_M=0.272$, $\Omega_\Lambda=0.728$, $h=0.704$, $\sigma_8=0.81$, $n_s=0.967$), not truncating its smallest-scale modes \citep{FalckEtal2017}. The 3D haloes we analyze have at least 100 particles, found with the Friends of Friends \citep[FOF,][]{DavisEtal1985} algorithm, using a standard linking length of 0.2 times the initial grid spacing. The $a=1/128$ initial conditions were generated with second-order Lagrangian perturbation theory. We give masses below in terms of the number of particles, of mass $m_{\rm part}=7.031\times 10^{10}M_\odot/h$.

\begin{figure}
    \begin{center}
      \includegraphics[width=\columnwidth]{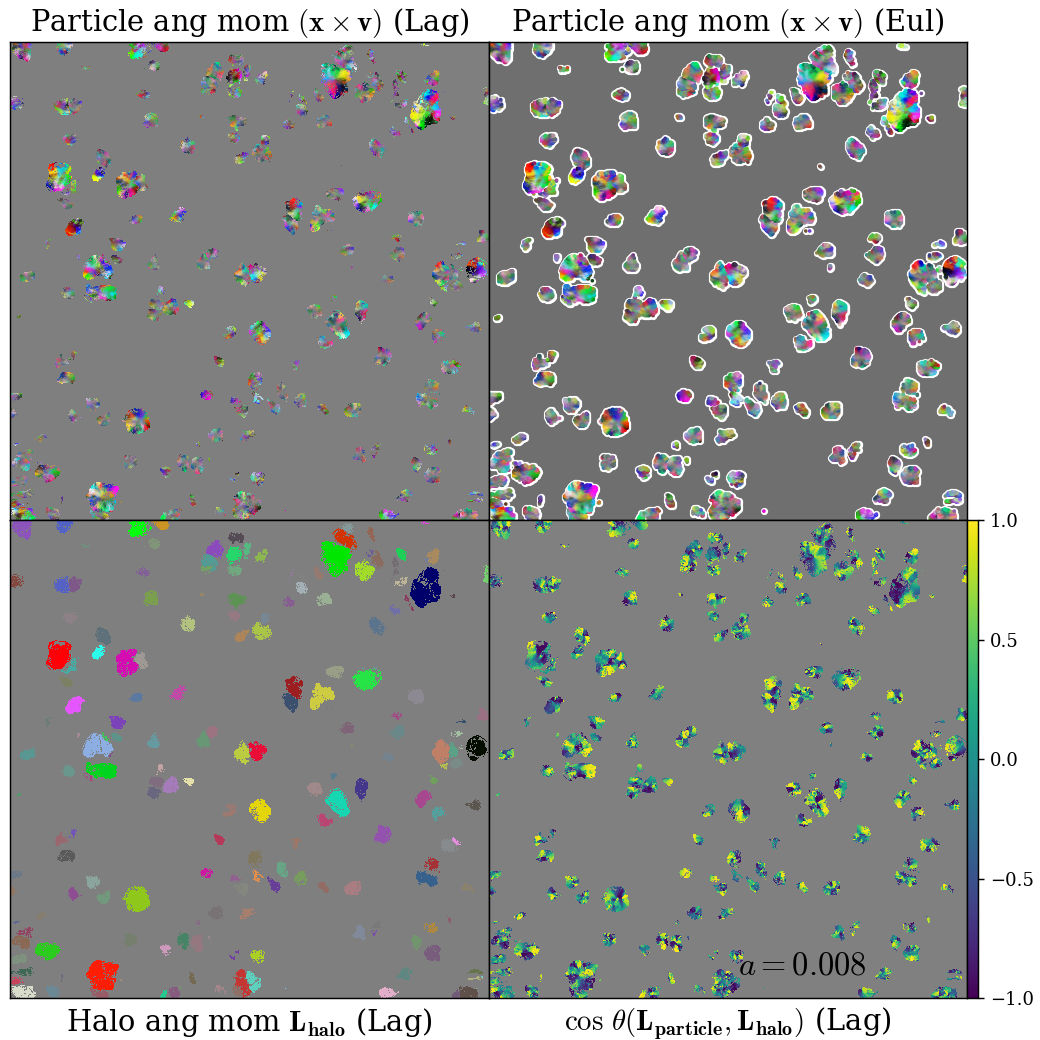}
    \end{center}  
    \caption{Quantities in the initial snapshot related to halo spin in a random patch, 480 particles (469 Mpc/$h$) on a side, of a 2D Lagrangian sheet (coplanar in the initial conditions). The patches are inside FOF haloes with at least 1000 particles in the final snapshot. The bottom-right panel shows a scalar: the dot product of the spin contributed by each particle with the total spin of its halo. In the other panels, colors show 3D vectors $\vec{L}$. Vectors with high magnitude $L$ have high or low luminance, and the hue indicates the direction.\footnote{The $(R,G,B)$ coordinates of color space are set according to $R=0.5+L_{\sinh} \sinh^{-1}(L_x/L_{\sinh})/L_{\rm max}$, with $L_{\sinh}$ set by hand, and $L_{\rm max}$ set to 3/4 of the maximum excursion from zero in any direction, in the initial snapshot. These values at top (left and right) and bottom-left are $(L_{\sinh},L_{\rm max})=(500, 1424)$ and $(10^5,4\times 10^5)$ km/sec $m_{\rm part}$ Mpc/$h$. $G$ and $B$ values come in the same way from $y$ and $z$ coordinates.} Grey pixels indicate a zero vector. Top panels show each particle's contribution to its halo's spin in Lagrangian and Eulerian coordinates, and the bottom-left panel shows its halo's spin. All angular momenta $L$ are normalized by $a^{3/2}$.  Animation at \url{https://www.youtube.com/watch?v=ZlxCbbiChTo&list=PLJxZSP5_6rBmo7Rz8JgnE1eI2bxRGXGCT}}
    \label{fig:crossy_bighalo_000}
%  \end{minipage}
%\end{figure*}
\end{figure}

\begin{figure}
%\begin{figure*}
%  \begin{minipage}{175mm}
    \begin{center}
      \includegraphics[width=\columnwidth]{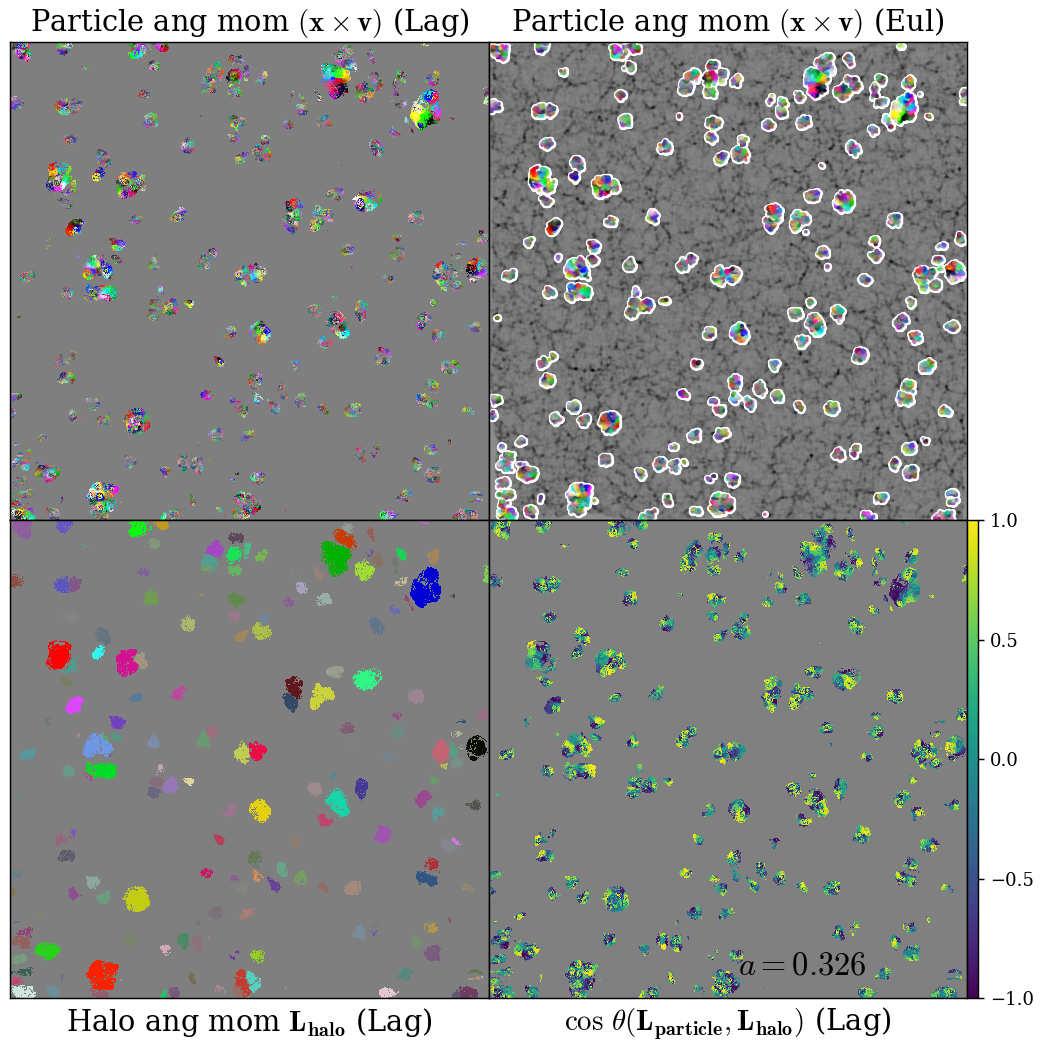}
    \end{center}  
    \caption{The same as Fig.\ \ref{fig:crossy_bighalo_000}, at $a=0.326$.}
    \label{fig:crossy_bighalo_032}
%  \end{minipage}
%\end{figure*}
\end{figure}

\begin{figure}
%\begin{figure*}
  %\begin{minipage}{175mm}
    \begin{center}
      \includegraphics[width=\columnwidth]{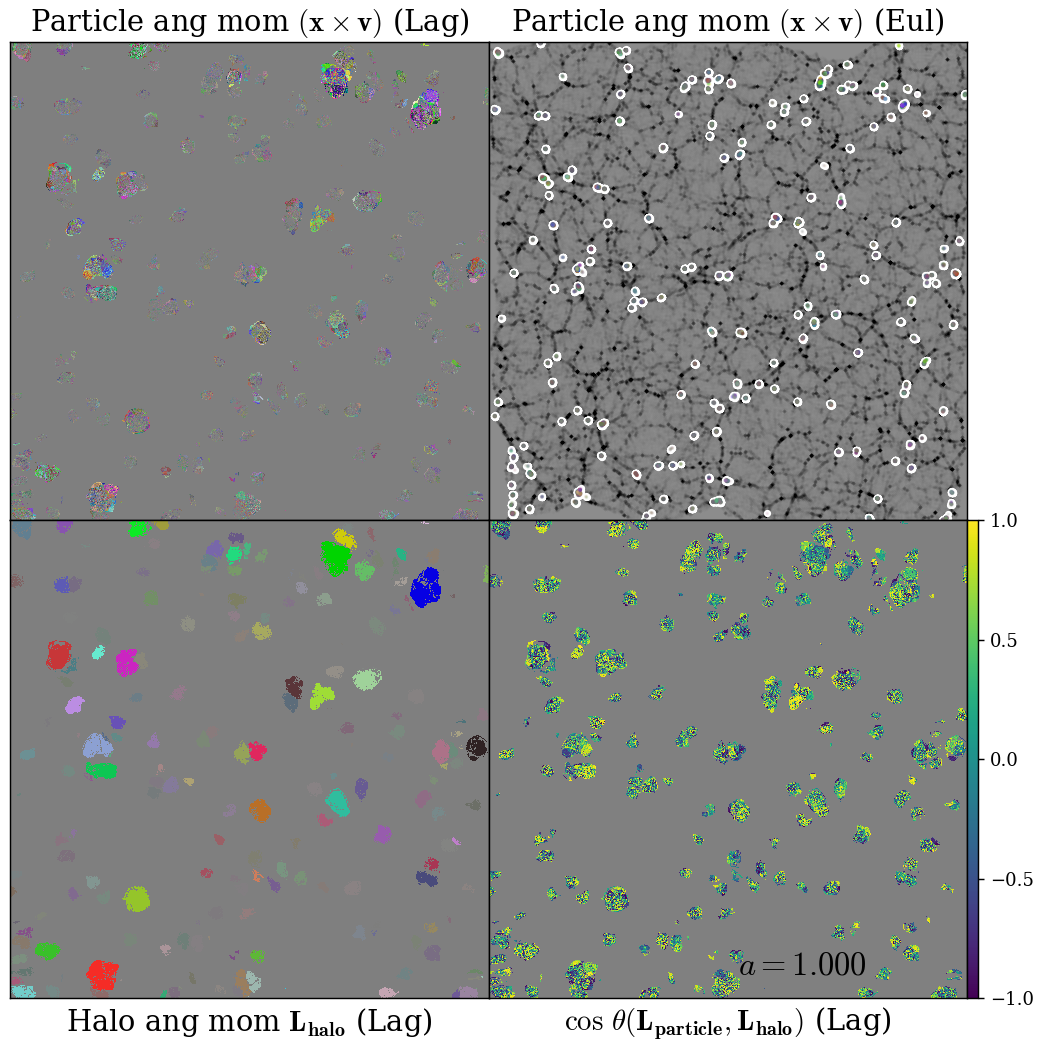}
    \end{center}  
    \caption{The same as Fig.\ \ref{fig:crossy_bighalo_000}, at the final snapshot.}
   \label{fig:crossy_bighalo_063}
%   \end{minipage}
%\end{figure*}
\end{figure}

Figs.\ \ref{fig:crossy_bighalo_000}-\ref{fig:crossy_bighalo_063} show 3D versions of information in Figs.\ \ref{fig:locangmom} and \ref{fig:spins_lageul}. Three of the panels show vector quantities, exploiting the 3D color space (with apologies to the colorblind). The precise quantities and colors being shown are not so important; we are mainly trying to show their persistence in time. One notable feature is that left-hand panels become more grey with time; this decay in a particle's contribution is also shown below in Fig.\ \ref{fig:particledots}.

As in the 2D case, a particle's contribution to its halo's eventual spin largely persists until it participates in halo collapse, after which it becomes rather random. Importantly for our conclusions, also particles' $\cos\theta$, where $\theta$ is the angle between a particle's contribution to its halo's spin, and its halo's total spin, is largely preserved with time, again until orbits become rather chaotic at late times.

%\begin{figure*}
%  \begin{minipage}{175mm}

Fig.\ \ref{fig:particledots} shows trajectories of $\vec{L}(a)\cdot \vec{L_{\rm init}}/(a^{3/2} L_{\rm init}^2)$ for random particles from a 1024$^2$ slice. $L$ here is a particle's contribution to its halo's spin. Fig.\ \ref{fig:particledots} also shows percentiles of these quantities over all particles in the slice, computed for each snapshot separately.

As shown at top, particles typically follow the median $L(a)\propto a^{3/2}$ curve at early times, but some of them veer systematically above or below that; this may be driven by different effective velocity growth rates in different regions. Eventually, most halo particles experience a spike in this quantity (as indicated by short dotted lines in the plot). This may happen when the particle actually enters the halo.

\begin{figure}
  \begin{center}
    \includegraphics[width=\columnwidth]{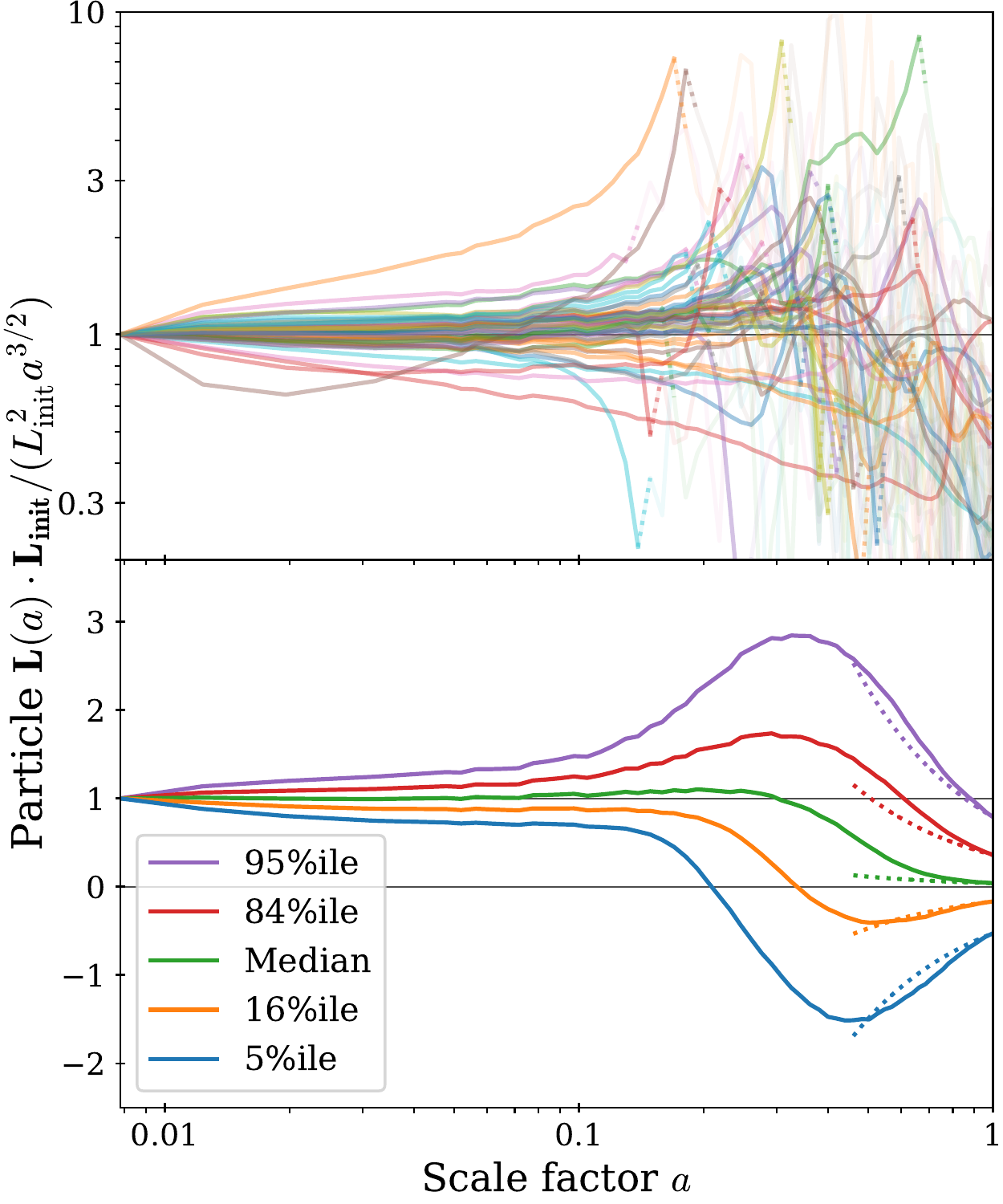}
  \end{center} 
  \caption{\textit{\textbf{Top}}: Trajectories in time of the contribution of random halo particles to their haloes' spins, normalized by their initial values. To indicate the transition from nearly linear-theory behavior to more chaotic motions, trajectories are shown with high opacity until the first sharp spike, then with a dotted line for one more snapshot, and thereafter with low opacity. \textit{\textbf{Bottom}}: Percentiles (corresponding to -2,-1,0,1, and 2$\sigma$ fluctuations in a Gaussian distribution), computed at each snapshot separately, of the same quantity. Dotted curves extrapolate $\propto a^{-3/2}$ backward from the last snapshot.
   }
  \label{fig:particledots}
\end{figure}

\begin{figure}
  \begin{center}
    \includegraphics[width=\columnwidth]{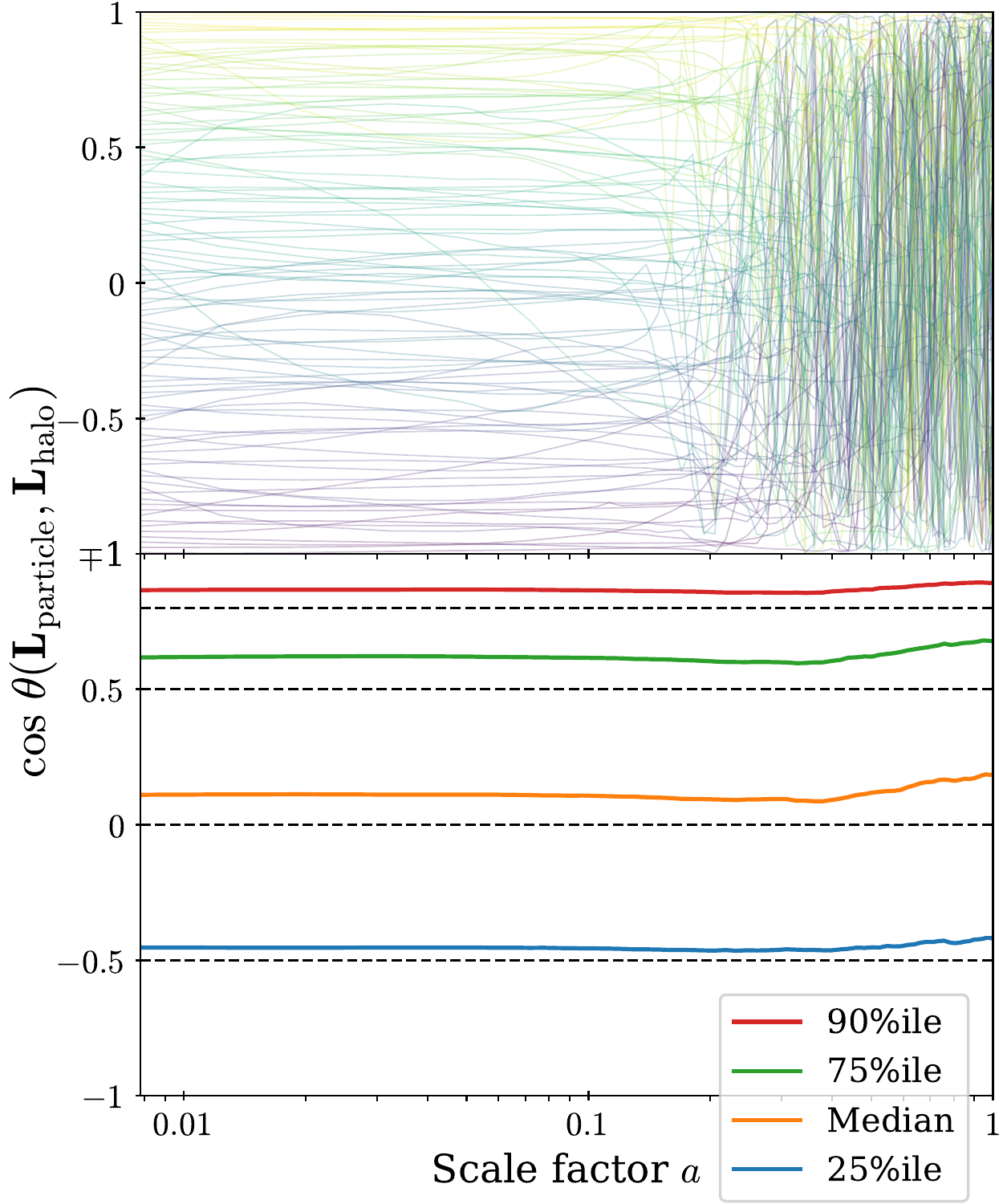}
  \end{center} 
  \caption{ \textit{\textbf{Top}}: Trajectories in time of the alignment of particles' spins with the total spins of their haloes. The trajectories, from random halo particles, begin equally spaced in the rank of the initial $\cos\theta_{\rm init}$, and remain colored according to that quantity. While there are wild oscillations at late times, note that the eye is drawn preferentially to these, and many particles behave more calmly, as well.  \textit{\textbf{Bottom}}: Percentiles of $\cos\theta$ with time. Dashed lines show where nearby colored curves would be if the distribution were entirely uniform. It is remarkable how small this deviation is, and how similar its distribution remains with time.}
  \label{fig:particlecosdist}
\end{figure}

Fig.\ \ref{fig:particlecosdist} shows the distribution of the alignment between particles' contribution to their haloes' spin with the total spin. In a literal interpretation of the TTT, in which particles in the protohalo torque up together, this quantity would be 1 for all particles. In the simulation, on the contrary, particles contribute very differently (even almost half negatively) to the halo's total spin. It is striking how close to uniform the distribution is, nearly random (in complete randomness, colored curves would coincide with the dashed lines), only ramping upward a bit at late times, perhaps as many haloes settle into their $z=0$ forms, as dynamical friction acts. As the top panel shows, many particles oscillate in this quantity, after they enter their haloes, but curiously, the total distribution of $\cos\theta$ changes only a bit.

A simple measure of the anisotropy of $\cos\theta$ is $\avg{\cos\theta}=0.07$ (initial), and 0.12 (final). Below, we will use this in units of the standard deviation of $\cos\theta$; $\avg{\cos\theta}/\sigma_{\cos\theta}=0.12$ (initial), and 0.19 (final).

\section{Halo spin in 3D}
\label{sec:halospin}
As shown in the previous section, the behavior of particles within haloes is not according to a naive, literal interpretation of the TTT. Nonetheless, many authors have found general agreement with its predictions, for whole haloes.

\begin{figure}
  \begin{center}
    \includegraphics[width=0.9\columnwidth]{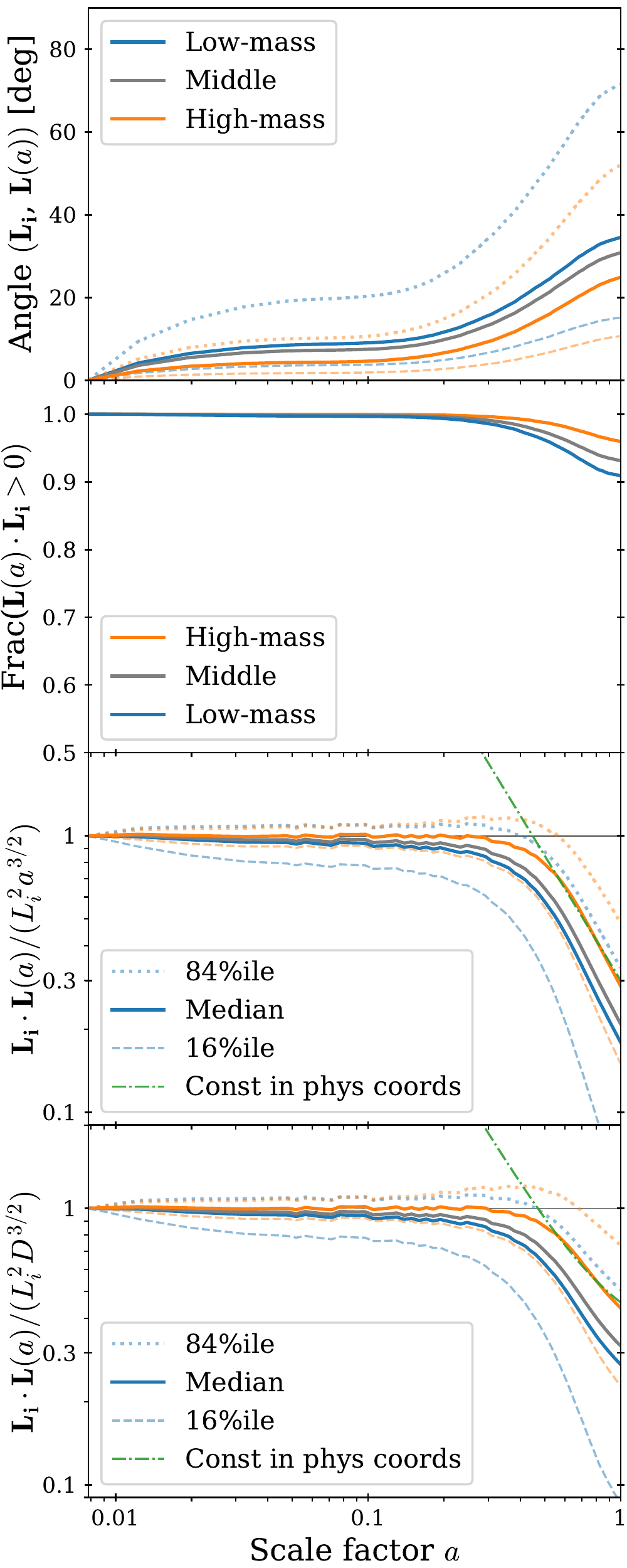}
  \end{center} 
  \caption{\textit{\textbf{Top}}: Deviation in angle between the initial spin vector and the spin at a later scale factor. It appears for three mass bins: the upper quartile, the two middle quartiles (`middle'), and the lower quartile. The dotted and dashed curves, noted in the middle-panel legend, show percentiles corresponding to $\pm 1\sigma$ in a Gaussian distribution. \textit{\textbf{Second}}: The fraction of haloes in each bin staying aligned with their initial protohaloes within $90^\circ$.  \textit{\textbf{Third}}: The median (and $\pm 1\sigma$ percentiles) scaling of halo spin with time, in units of its initial value. It grows as $a^{3/2}$ (constant, here) until $a\approx 0.4$, and then becomes constant with time ($\propto a^{-3/2}$ here). For clarity, the $\pm 1\sigma$ percentiles are only shown for the high- and low-mass bins.  \textit{\textbf{Bottom}}: The same, normalizing by $D^{3/2}$ instead of $a^{3/2}$.}
  \label{fig:indra_dots}
\end{figure}

Fig.\ \ref{fig:indra_dots} shows several plots that are broadly in agreement with these results in the Indra simulation, for three halo bins. Low-, middle-, and high-mass bins include haloes in the bottom, middle two, and top mass quartiles; the bin edges are at 100, 132, 361, and 46383 particles, i.e.\ masses of $7.0\times 10^{12}$, $9.3\times10^{12}$, $2.5\times 10^{13}$, and $3.3\times 10^{15}$ $M_\odot/h$ (the highest halo mass in the box).

As shown in the top panel, the haloes in our sample remain somewhat aligned with their initial protohalo's spin, the middle-mass sample growing misaligned by $\sim 30^\circ$ at $z=0$. P02 found more misalignment using the ZA, typically $\sim 40^\circ$ at $z=0$, likely because of the smaller haloes analyzed there. The second panel shows the fraction of haloes in each sample flipping their spin directions by $>90^\circ$ compared to the initial conditions; less than 10\% of them flip at $a=1$, even in the low-mass sample.

The third panel addresses spin amplitudes, as well, and shows (with much better statistics) essentially what was shown by W84. In the median, spins evolve as though intra-halo velocities grow with linear theory, until a characteristic time after which the spin is roughly conserved in physical coordinates. Although we call this the collapse time, we have not actually checked that haloes tend to `collapse' (a word subject to definition details) then. It would be interesting to compare the time at which individual haloes depart from the linear-theory trajectory with other definitions of collapse.

Note that small haloes begin their constant-$L$ phase (decay, in the middle panel of Fig.\ \ref{fig:indra_dots}) earlier than large haloes. This makes sense; large haloes typically collapse later, and have larger spin, so that initial spin persists longer.  Secondary interactions and angular momentum exchanges also contribute to this downturn; these preferentially affect smaller haloes. These effects may be related to the slightly steeper than $L=$const slope in the median.

While by default we assume a $L\sim a^{3/2}$ scaling, most relevant for a matter-dominated universe, we also test a scaling with the growth factor $D$ rather than the scale factor, $L\sim D^{3/2}$. The downturns in $L$ coincide roughly with the epoch where dark energy starts to be substantial, and $D$ departs from $a$. The last panel shows the result of normalizing by $D^{3/2}$ instead of $a^{3/2}$. The curves do not change qualitatively, but using $D$ does extend the regime where the curves are nearly 1, and softens the downturns.

\subsection{Comparison to the Zel'dovich Approximation}
\label{sec:halospinzeld}
The full ZA has been tested before as a framework to predict primary halo spin (i.e.\ until collapse, and secondary encounters). The ZA was found generally to be superior to the TTT estimate (P02). Here we additionally test the SPIM prescription, of estimating the spin only from initial velocity field in the protohalo and scaling that by $a^{3/2}$. Velocities grow identically in SPIM and the ZA, as $a^{1/2}$. The only difference is that in SPIM, particles stay in their initial Lagrangian positions. This ends up giving only a tiny difference numerically, but the SPIM prescription is much simpler; given a protohalo, it requires only a measurement in the initial conditions. The ZA, on the other hand, requires more computation, and does not offer a strong intuitive understanding in this context of halo spin.

\begin{figure}
 \begin{center}
 \includegraphics[width=\columnwidth]{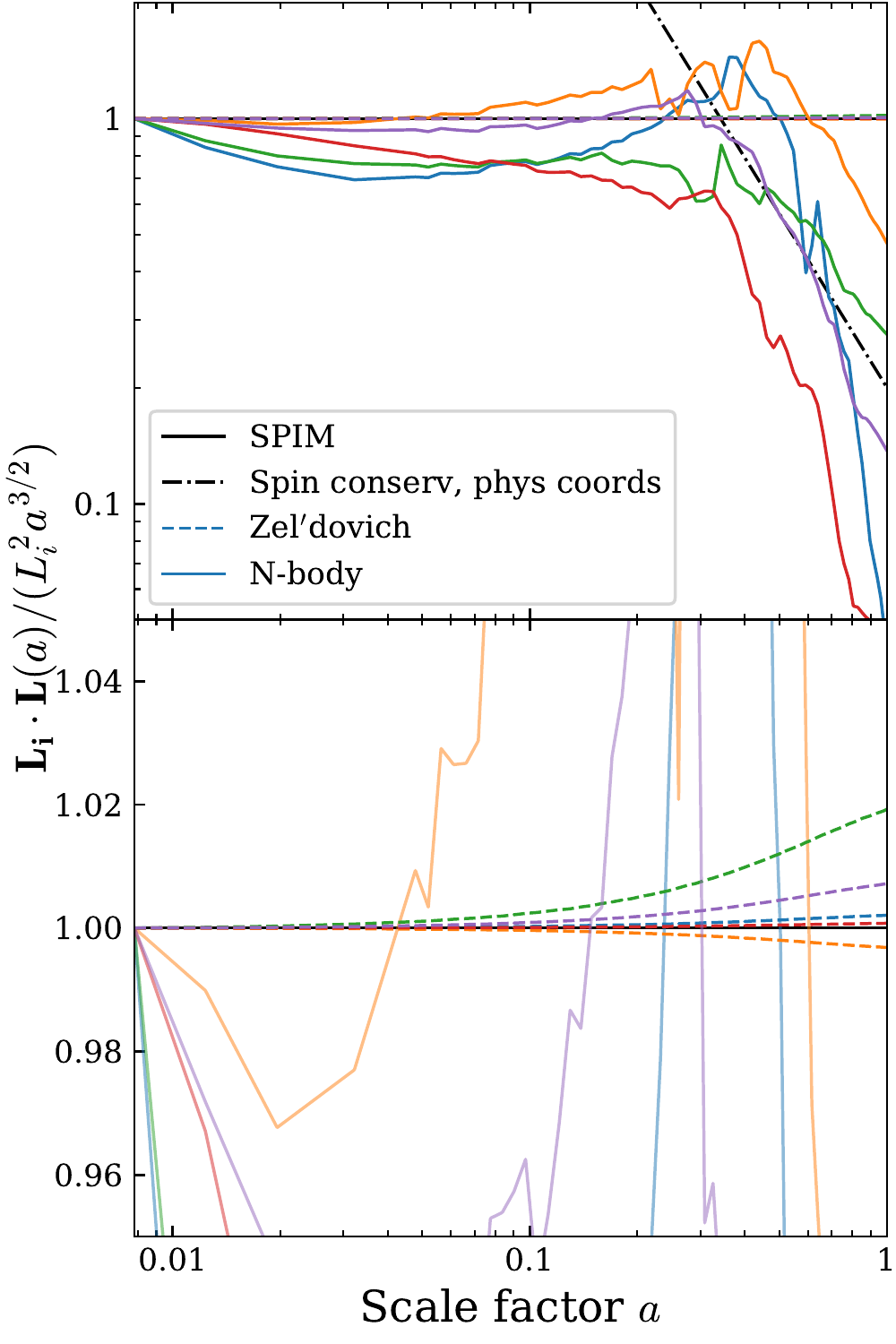}
 \end{center}
 \caption{Spin trajectories with time (solid colored curves), compared to the SPIM prediction of conservation of linear-theory angular momentum (horizontal black line at 1), for the most-massive five haloes in the Indra simulation. The Zel'dovich (ZA) predictions are shown in the same colors, with dashed curves.  Once haloes collapse, $\vec{L}$ remains roughly conserved, scaling as the dot-dashed black line $\propto a^{-3/2}$. The bottom panel is zoomed in, to clarify how the ZA and SPIM differ.}
  \label{fig:zeldovichcomp}
\end{figure}

Fig.\ \ref{fig:zeldovichcomp} shows trajectories of $N$-body halo spins compared to their SPIM and ZA predictions, for the five biggest haloes in the above Indra simulation, plotted as in Fig.\ \ref{fig:indra_dots}. The ZA and SPIM predictions are nearly the same, essentially indistinguishable before zooming in, shown in the lower panel. For the haloes shown, the ZA prediction even typically veers in the opposite direction than $N$-body at low redshift, indicating that SPIM is actually {\it more accurate} than ZA (however, with a negligible improvement compared to the total error). While the ZA can be surprisingly effective at predicting the cosmic web, individual particle positions in the ZA are known to grow quite inaccurate eventually; it is plausible that it is more accurate simply to leave particles at their initial comoving positions. In any case, it seems that for the purpose of spin estimation, keeping track of particle displacements in the ZA seems not worthwhile compared to SPIM. 

\section{Scaling of spin with mass from a correlated random walk}
\label{sec:randomwalk}
Does the SPIM concept help to understand any other fundamental properties of halo spin? One key result, as many previous papers \citep[e.g.][]{EfstathiouJones1979,HeavensPeacock1988,CatelanTheuns1996} have found, is that for a finally collapsed halo, typically $L\propto M^{5/3}$, where $M$ is the total halo mass. It may be a coincidence, but this relationship even seems to hold within the errors for total stellar mass and angular momentum across a wide range of galaxies \citep{FallRomanowsky2018}.

\begin{figure}
  \begin{center}
    \includegraphics[width=\columnwidth]{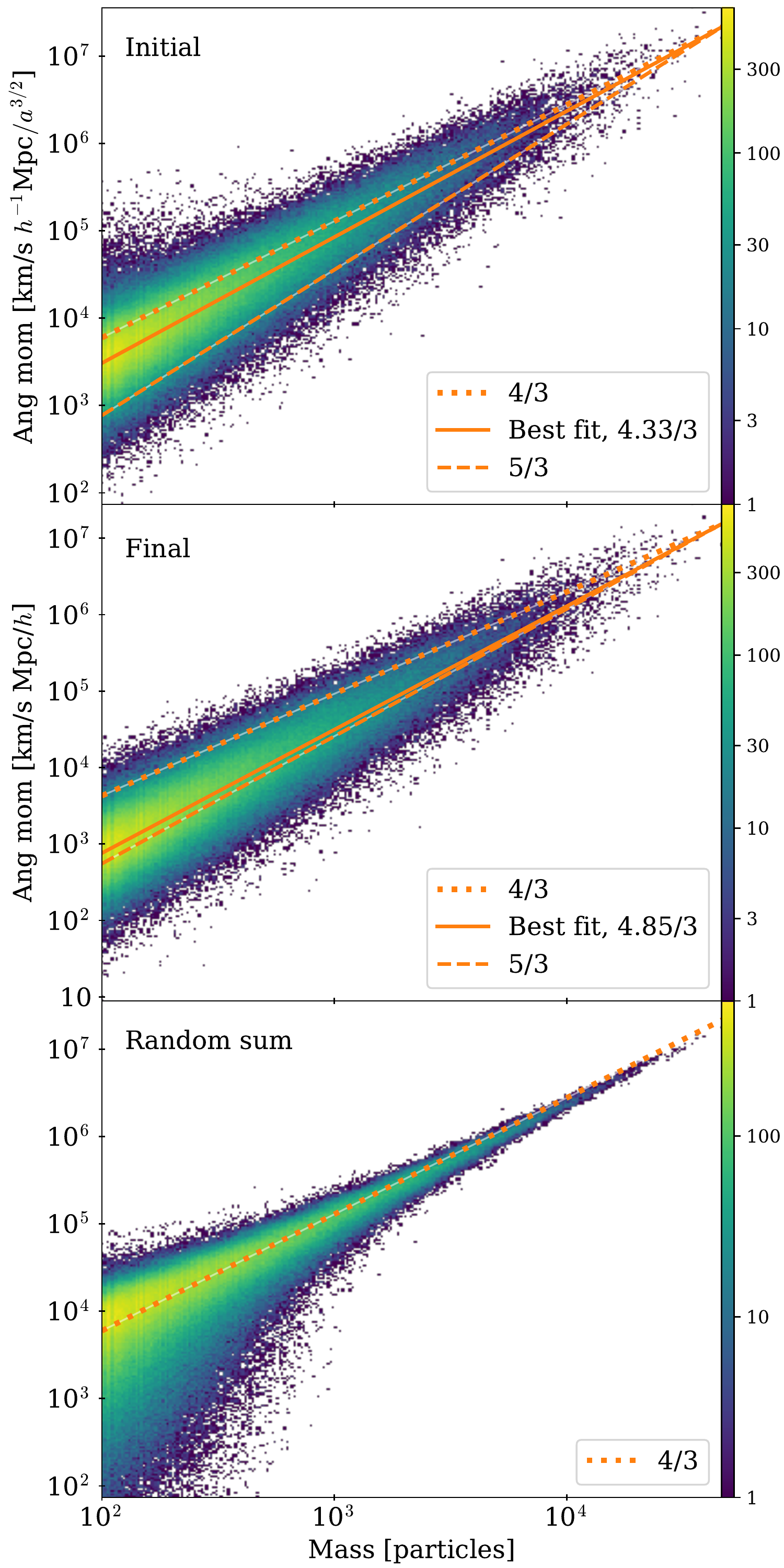}
  \end{center} 
  \caption{2D histograms of halo mass $M$ vs.\ spin $L$, for initial protohaloes and final haloes (first two panels). In calculating $L$, particles were considered to have unit mass, as indicated in the $x$-axis label. Note the factor of $a^{3/2}$ in the top panel; with this factor, if all haloes evolved according to linear theory, the top and middle panels would be identical. We also show the results of a toy model of a correlated random walk (see text), with each particle contributing one random variate to its halo in its initial conditions. The color (with logarithmic colorbar at right) shows the number of haloes in a pixel of the histogram. The orange lines are of the form $L\propto M^{\alpha}$, tied to coincide for the highest mass in the sample; the legend shows $\alpha$.}
  \label{fig:mass_am}
\end{figure}

As Fig.\ \ref{fig:mass_am} shows, we also find that this relationship holds rather well for final haloes in the simulation. The best fit has $L\propto M^{4.85/3}$, a simple least-squares linear fit in log space. But the relationship is different for initial protohaloes, which typically have a shallower slope, closer to $L\propto M^{4/3}$ (with best fit $M^{4.33/3}$).

Why the change in slope from the initial to final conditions? It is because $L/a^{3/2}$ stays almost constant for large haloes, but decays for small haloes. The largest haloes stay above $10^7$, but for the smallest haloes, the best-fitting line goes down by about a factor of 3 from initial to final conditions. This is as in the middle panel of Fig.\ \ref{fig:indra_dots}: large haloes tend to retain their initial spin longer than small haloes do.

Consider a toy model of initial protohalo angular momentum as coming from a correlated random velocity field; this will help us make sense of the scaling found in the initial conditions. We choose the $z$-axis to align with the total spin, and $v_\phi$ to be the azimuthal component of the velocity.
\begin{align}
L_z & = \rho_0\int_{V_L} (\vec{r}\times \vec{v})\cdot \vhat{z} 
 dr d\Omega\nonumber\\
 & \propto \int_{V_L} r v_\phi r^2 dr d\Omega.
\label{eqn:lz1}
\end{align}

In the completely correlated extreme that $v_\phi$ is a constant,
\begin{equation}
L_z\propto \int_{V_L}r^3 dr d\Omega\propto R^{4},
\end{equation}
where $R$ is an effective radius of the (nearly spherical) protohalo, giving $L\propto M^{4/3}$.

In the opposite extreme, $v_\phi$ is an uncorrelated Gaussian random number at each particle. The angular momentum is a sum over particles of $r v_\phi$. The contribution $L_z(r)dr$ from a shell at radius $r$ and thickness $dr$ will be a sum over particles, $r\sum 4\pi r^2 n v_r dr$, where $n$ is the number density of particles. If at each particle, $v_\phi \sim G(0,\sigma_v)$, where $\sigma_v$ is the dispersion of a single component of the initial velocity, $L_z(r)dr$ will be the expected displacement of a random walk, i.e.\ the square root of the number of random contributions, $\propto r dr$. In that case, the power of $r$ is reduced by one in the integrand of Eq.\ (\ref{eqn:lz1}), so $L_z\propto r^3$, and $L\propto M$. We checked this scaling numerically by recomputing halo spins as in the top panel of Fig.\ \ref{fig:mass_am}, and setting $v_\phi$ at each particle to an uncorrelated Gaussian random number. This is not the observed behavior.

The reality can be approximated to be between these extremes; a correlated random walk. To crudely approximate the effect of velocity correlations, let us assume that $v_\phi \sim G(0.12\sigma_v,\sigma_v)$, i.e. with a mean 0.12 times the standard deviation, as measured above from the distribution of $\cos\theta$, the alignment of the total halo spin to each particle's contribution. ($\cos\theta$ is a uniform, not Gaussian variate, but the Gaussian-distributed velocity gets multiplied by this.)

For the bottom panel of Fig.\ \ref{fig:mass_am}, we kept fixed the positions of all particles in protohaloes to be the same as in the actual simulation, but set $v_\phi$ at each particle according to $v_\phi\sim G(0.12\sigma_v, \sigma_v)$. We set $\sigma_v=180/a^{1/2}$ km/sec, the average of $\sigma(v_x)$, $\sigma(v_y)$ and $\sigma(v_z)$ for all particles in the initial conditions.

As this shows, setting $\avg{v_\phi}/\sigma_v=0.12$ is enough to give this relation, $L\propto M^{4/3}$, as well. The scatter about the mean is inaccurate, however; a more accurate treatment of the distribution of $v_\phi$ is needed for this. Notice that the dispersion in the top and bottom panels looks rather comparable at a mass of $1000$ particles, the lower mass limit of the measurement used to estimate $\sigma_v/\avg{v_\phi}$. If this quantity were measured separately for each mass, we would expect the scatter to match better in the top and bottom panels.

\section{Rotation from an irrotational flow}
\label{sec:kelvin}
Before concluding, in the attempted pedagogical spirit of this paper, it is helpful to clarify (elaborating on a short comment by \citet{Doroshkevich1970}) how something could come to rotate from `random,' irrotational motions. Judging from the words without their precise physical meaning, it seems impossible. But `irrotational' means zero-vorticity, which pertains to infinitesimal volume elements, whereas rotation (in the sense of angular momentum or spin) is an integrated quantity.

Still, rotation requires special conditions to appear in an irrotational flow. A {\it spherical} patch of an irrotational, homogeneous-density velocity field must have vanishing spin. Returning to Eq.\ \ref{eqn:lz1}, for a spherical patch of radius $R$, putting the origin at the center of the sphere, and choosing an arbitrary direction $\hat{z}$,
\begin{align}
L_z&= \rho_0 \int_0^\pi \int_0^\R \left(\int_0^{2\pi} rv_\phi d\phi\right) r^2 dr \sin\theta d\theta\\
&= \rho_0 \int_0^\pi \int_0^R C(r,\theta) r^2 dr \sin\theta d\theta,
\end{align}
where $C(r,\theta)=\oint_{(r,\theta)}\vec{v}\cdot d\vec{s}$ is the circulation in the circular ring at radius $r$ and altitude $\theta$. By Kelvin's circulation theorem, since the vorticity is zero inside the ring, $C(r,\theta)=0$, so $L_z$ = 0.

Note that this argument that $L_z=0$ in an irrotational flow depends on circulation paths being circular and azimuthal; otherwise, the component of $\vec{v}$ summed up over the path would not be purely $v_\phi$. Also, $L_z=0$ in an irrotational flow for some aspherical shapes; the argument applies to any cylindrically symmetric shape that is a sum of azimuthal rings, e.g.\ a cylindrical filament, an ellipsoid, or even a torus. However, for an object with only one axis of cylindrical symmetry, the argument only applies for the component of $\vec{L}$ along that axis. To ensure $\vec{L}=\vec{0}$, the shape has to have spherical symmetry.

\begin{figure}
  \begin{center}
    \includegraphics[width=0.8\columnwidth]{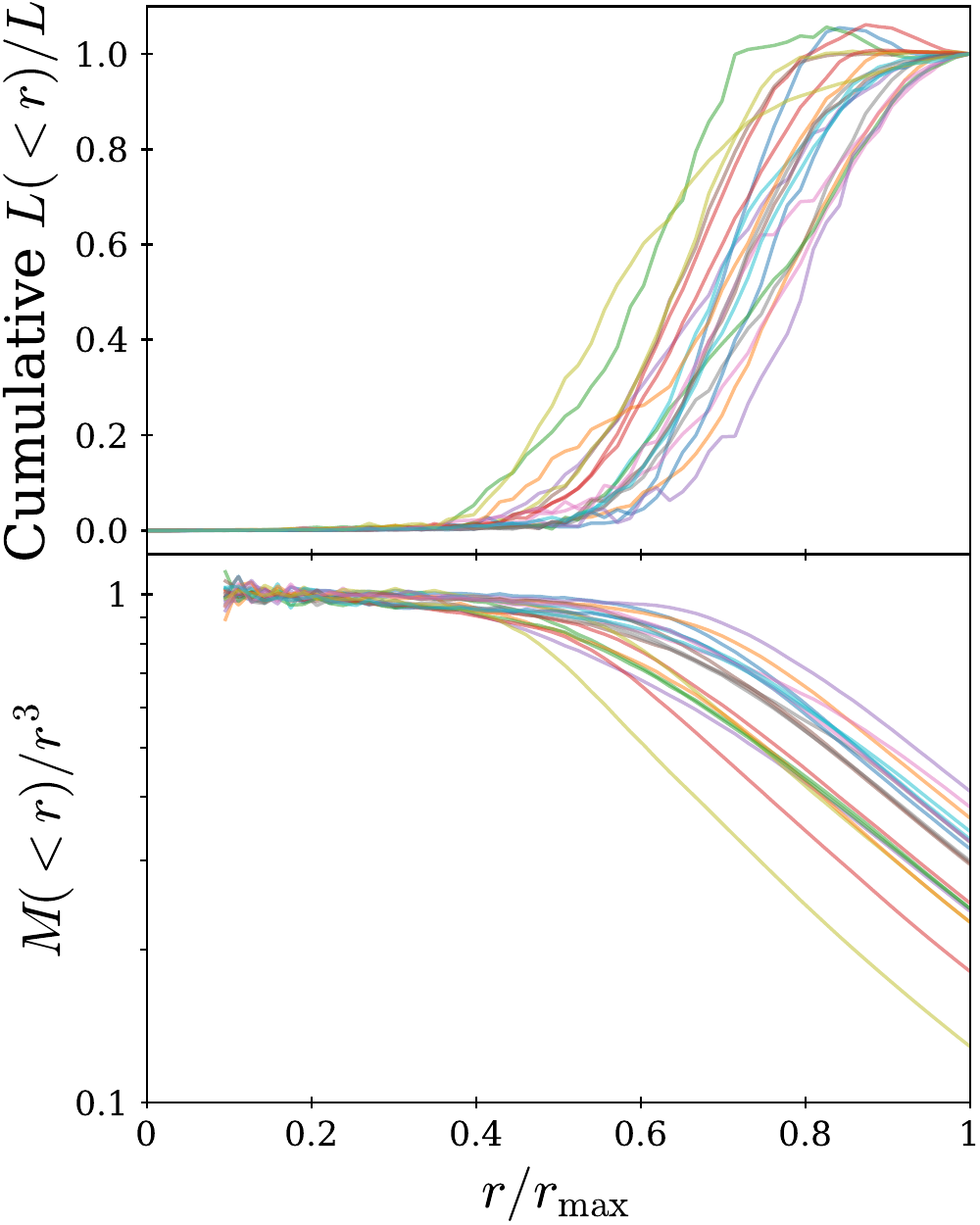}
  \end{center} 
  \caption{\textit{\textbf{Top}}: Magnitudes of protohalo (initial-conditions) angular momenta, summed cumulatively in spherical shells from protohalo volume centroids, for the 20 most massive haloes in the simulation. These are normalized by the total angular momentum magnitude, tying all curves at $1$ at $r=r_{\rm max}$, the distance to the farthest particle. \textit{\textbf{Bottom}}: The mass inside radius $r$ divided by $r^3$ for these 20 haloes, colored the same as at top, normalized to be $\sim 1$ for low $r$. The curves are horizontal until $r/r_{\rm max}\sim 0.4$-$0.6$, but turn down after that, meaning that the haloes are filled spheres inside this radius, but become aspherical thereafter. Curves at top turn up at about the same radius where curves at bottom turn down.}
  \label{fig:cumulativespin}
\end{figure}

For an aspherical patch, there can be contributions to the angular momentum from outside a maximal sphere that fits in the patch, with center at the volume centroid. Fig.\ \ref{fig:cumulativespin} shows this, for the most massive haloes in our 3D simulation. The angular momentum, cumulatively added in spherical shells of center at protohalo volume centroids, is zero (except for some small noise) as long as $M(r)\propto r^3$, i.e.\ inside a maximal sphere. But the magnitude $L>0$ outside this region.

\begin{figure}
  \begin{center}
    \includegraphics[width=\columnwidth]{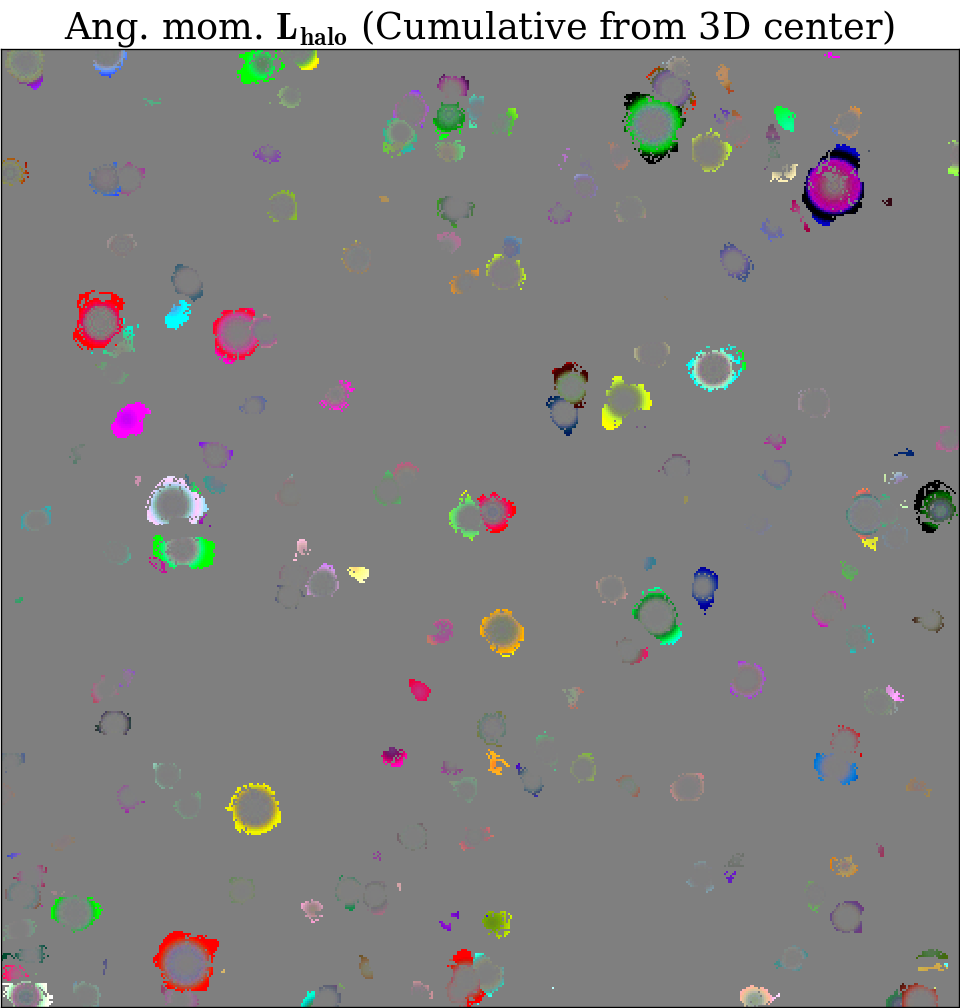}
  \end{center} 
  \caption{The bottom-left panel of Fig.\ \ref{fig:crossy_bighalo_000}, except summing particles' contribution to protohaloes' angular momenta cumulatively, starting from the center. Note the grey centers of almost all haloes. Each protohalo has a maximal grey sphere, centered at its centroid, that can fit in the protohalo. The total contribution to halo angular momentum within this sphere is zero, by Kelvin's circulation theorem. But regions outside this sphere contribute. Note that a halo center in this slice may not be grey if the 2D slice is far from the halo's 3D center.}
  \label{fig:cumulative_spin_000}
\end{figure}

Fig.\ \ref{fig:cumulative_spin_000} shows the same, in another way; it shows the same 2D patch from Fig.\ \ref{fig:crossy_bighalo_000}, with spin cumulatively added from the center of each halo. The key thing to note is that the regions inside maximal spheres are grey, i.e.\ have zero cumulative spin.

This recalls the TTT, in which a spherical patch, with vanishing moment of inertia, cannot spin up. This is why non-spherical, ellipsoidal patches are considered in the TTT. But again, in a naive literal interpretation of the TTT, one would expect an entire ellipsoid to be torqued up; what actually happens is that the velocity field inside the protohalo, but outside a maximal sphere around the volume centroid of an aspherical protohalo, determines the spin.

This interpretation also sheds light on why the angular momentum of a growing halo (keeping track of only the bound part of the halo, not all the matter that ends up in the final halo) can fluctuate wildly with time \citep[\eg][]{ContrerasEtal2017}. It is obvious that as haloes grow, newly accreted matter contributes with some randomness to their angular momentum. But not only that; it is {\it only} the last-accreted outskirts of the halo, outside the maximal Lagrangian sphere, that contribute to its net linear-theory angular momentum. Inside the maximal Lagrangian sphere, the net angular momentum is zero.

This maximal-sphere finding is related to results in our previous sections. For example, within the maximal sphere of a protohalo in the initial conditions, the distribution of $\cos\theta$ in Fig.\ \ref{fig:particlecosdist} is likely consistent with uniform; the excess $\avg{\cos\theta}=0.12$ coming only from regions outside the protohalo's maximal sphere.

This discussion is also relevant to wave dark matter ($\psi$DM) \citep[\eg][]{SchiveEtal2014,Marsh2016}, because the dark-matter velocity field in this model must be irrotational at all redshifts (since the velocity field is defined via a gradient, and therefore has no curl). \citet{DuEtal2018} found that a $\psi$DM solitonic core in a subhalo resembles an irrotational Riemann S-ellipsoid \citep[see also][]{Rindler2012}. Any spherically symmetric patch of this ellipsoid has zero angular momentum. But the cores do rotate, in the sense that their boundaries are aspherical, and have nonzero angular momentum summed up to those boundaries.

\section{Conclusions}
We describe a SPIM (Spin from Primordial Inner Motions) picture of how dark-matter haloes come to spin. It uses the velocity field, sourced gravitationally from the pattern of density fluctuations within and around the protohalo. The prescription is as follows: measure the angular momentum imprinted onto the initial velocity field within the protohalo, and propagate that according to linear theory, $\vec{L}\propto a^{3/2}$, until collapse. After collapse, treat $\vec{L}$ as constant. We leave vague what `collapse' physically means, besides the turning point in this plot.

We hope that SPIM will be helpful intuitively and conceptually, as a complement to the usual explanation, the tidal torque theory (TTT). A possible misconception in the TTT, for someone first learning about it, might be that protohaloes are literally ellipsoids, clearly demarcated {\it a priori} in the initial conditions, that uniformly torque up. Instead, the boundaries are not clear in the initial conditions, but in a simulation, can be traced back in time from the matter that has collapsed in the final conditions. The velocity field within each halo retains much dispersion, at all times, but generally carries net angular momentum, coming from the velocity field in its outskirts. The initial velocity field is a lower-order quantity than the tidal field, relatively easy to visualize. We find the idea of conservation (and linear-theory growth) of angular momentum within a collapsing body to be more familiar and intuitive than considering the tidal field, smoothed differently based on the scale of the protohalo, and its misalignment with the inertia tensor.

There are some modest practical quantitative advantages of SPIM over the TTT, as well. SPIM gives almost the same results as the ZA (the Zel'dovich approximation, more accurate than the TTT), but is simpler.  We also show how this picture can be used to understand the scaling of protohalo spin with mass, in terms of a correlated random walk.

The TTT, on the other hand, is more useful for analytic estimates and statistics of spin, since it does not rely on a detailed knowledge of the high-resolution velocity field and protohalo boundaries for each halo, as SPIM does. Given only the initial conditions of a simulation, and some approximate prescription to predict protohalo boundaries, the TTT may provide an estimate of halo spin of comparable accuracy to using SPIM based on these approximate boundaries.

However, both SPIM and the TTT are usefully applicable only to predicting `primary' halo spin, evolving before secondary encounters that may exchange angular momentum with structures that have not merged with the halo by the final conditions. It is only for the highest-mass haloes that this prediction is accurate for individual haloes. Lower-mass protohalo spins fluctuate substantially with time, and do not individually follow the SPIM prediction. However, the median behavior of even lower-mass haloes follows the prescription to an interesting degree.

This prescription can be applied to filaments, as well. We analyze a 2D simulation, in which the `haloes' correspond at some level to cross-sections of 3D filaments. As is perhaps not widely appreciated, filaments do generically spin, according to this analysis. Filamentary (2D halo) spin grows with $a^{3/2}$, just like 3D haloes. However, to conclude that filaments actually coherently spin, a full 3D analysis is required.

Also, connecting halo spin directly to the velocity field, instead of the tidal field, may make spins easier to use to infer initial conditions from observations, as in Bayesian reconstruction methods \citep[\eg][]{KitauraEnsslin2008,JascheLavaux2015}. Clusters spin particularly faithfully compared to the initial conditions, and occupy large volumes of Lagrangian space. It might be useful to include these constraints on the velocity field.

Also, unlike the TTT, SPIM would be an ideal formalism to estimate the contribution to halo spin from any (unexpected) initial vorticity.  Initial vorticity is usually thought not to have survived inflation, since a Fourier-space analysis predicts it to decay from expansion, in linear theory. Still, it would be useful to falsify that vorticity was completely negligible on small scales. In collapsing patches, it may be that initial vorticity would have survived. Suppose that the vortical part of the initial velocity field was nonzero, but still small enough to negligibly affect protohalo boundaries. The spin from irrotational motions cancels out within the protohalo's maximal sphere, but not the spin from vortical motions.  SPIM would predict that even a modest amount of vorticity inside the sphere could add up to substantially affect the final spin of a large halo, interesting to test in an $N$-body simulation. If primordial vorticity might be hiding in haloes and filaments, having expanded away elsewhere, SPIM would provide a framework to understand it.

\begin{acknowledgments}
MCN thanks Bernard Jones, St\'{e}phane Colombi, Jens Niemeyer, Qianli Xia and Cai Yanchuan for helpful and encouraging discussions, and Mike Fall for helpful comments. MCN is grateful for funding from Basque Government grant IT956-16. MAAC received support from Mexican grant DGAPA-PAPIIT IA104818, and JW was supported by the Chinese National Key Program (grants 2015CB857005, 2017YFB0203300) and NFSC grant 11873051. The `Vorticity in the Universe' workshop at the Aspen Center for Physics (supported by U.S.\ National Science Foundation grant PHY-1607611) sparked initial thoughts about this project.\\

MCN is responsible for the bulk of this article's content, with the essential contribution of other authors' expertise, data, and suggestions.
\end{acknowledgments}

\bibliographystyle{hapj}
\bibliography{refs}

\begin{thebibliography}{59}
\expandafter\ifx\csname natexlab\endcsname\relax\def\natexlab#1{#1}\fi

\bibitem[{{Arag{\'o}n-Calvo} {et~al.}(2007){Arag{\'o}n-Calvo}, {van de
  Weygaert}, {Jones}, \& {van der Hulst}}]{AragonCalvoEtal2007}
{Arag{\'o}n-Calvo}, M.~A., {van de Weygaert}, R., {Jones}, B.~J.~T., \& {van
  der Hulst}, J.~M. 2007, \apjl, 655, L5, astro-ph/0610249

\bibitem[{{Aragon-Calvo} \& {Yang}(2014)}]{AragonCalvoYang2014}
{Aragon-Calvo}, M.~A., \& {Yang}, L.~F. 2014, \mnras, 440, L46, 1303.1590

\bibitem[{{Bardeen} {et~al.}(1986){Bardeen}, {Bond}, {Kaiser}, \&
  {Szalay}}]{BBKS}
{Bardeen}, J.~M., {Bond}, J.~R., {Kaiser}, N., \& {Szalay}, A.~S. 1986, \apj,
  304, 15

\bibitem[{{Catelan} \& {Theuns}(1996)}]{CatelanTheuns1996}
{Catelan}, P., \& {Theuns}, T. 1996, \mnras, 282, 436, astro-ph/9604077

\bibitem[{{Codis} {et~al.}(2018){Codis}, {Jindal}, {Chisari}, {Vibert},
  {Dubois}, {Pichon}, \& {Devriendt}}]{CodisEtal2018}
{Codis}, S., {Jindal}, A., {Chisari}, N.~E., {Vibert}, D., {Dubois}, Y.,
  {Pichon}, C., \& {Devriendt}, J. 2018, \mnras, 481, 4753, 1809.06212

\bibitem[{{Codis} {et~al.}(2012){Codis}, {Pichon}, {Devriendt}, {Slyz},
  {Pogosyan}, {Dubois}, \& {Sousbie}}]{CodisEtal2012}
{Codis}, S., {Pichon}, C., {Devriendt}, J., {Slyz}, A., {Pogosyan}, D.,
  {Dubois}, Y., \& {Sousbie}, T. 2012, \mnras, 427, 3320, 1201.5794

\bibitem[{{Codis} {et~al.}(2015){Codis}, {Pichon}, \&
  {Pogosyan}}]{CodisEtal2015}
{Codis}, S., {Pichon}, C., \& {Pogosyan}, D. 2015, \mnras, 452, 3369,
  1504.06073

\bibitem[{{Contreras} {et~al.}(2017){Contreras}, {Padilla}, \&
  {Lagos}}]{ContrerasEtal2017}
{Contreras}, S., {Padilla}, N., \& {Lagos}, C.~D.~P. 2017, \mnras, 472, 4992,
  1705.03463

\bibitem[{{Davis} {et~al.}(1985){Davis}, {Efstathiou}, {Frenk}, \&
  {White}}]{DavisEtal1985}
{Davis}, M., {Efstathiou}, G., {Frenk}, C.~S., \& {White}, S.~D.~M. 1985, \apj,
  292, 371

\bibitem[{{Doroshkevich}(1970)}]{Doroshkevich1970}
{Doroshkevich}, A.~G. 1970, Astrophysics, 6, 320

\bibitem[{{Du} {et~al.}(2018){Du}, {Schwabe}, {Niemeyer}, \&
  {B{\"u}rger}}]{DuEtal2018}
{Du}, X., {Schwabe}, B., {Niemeyer}, J.~C., \& {B{\"u}rger}, D. 2018, \prd, 97,
  063507, 1801.04864

\bibitem[{{Efstathiou} \& {Jones}(1979)}]{EfstathiouJones1979}
{Efstathiou}, G., \& {Jones}, B.~J.~T. 1979, \mnras, 186, 133

\bibitem[{{Efstathiou} \& {Jones}(1980)}]{EfstathiouJones1980}
------. 1980, Comments on Astrophysics, 8, 169

\bibitem[{{Falck} {et~al.}(2017){Falck}, {McCullagh}, {Neyrinck}, {Wang}, \&
  {Szalay}}]{FalckEtal2017}
{Falck}, B., {McCullagh}, N., {Neyrinck}, M.~C., {Wang}, J., \& {Szalay}, A.~S.
  2017, \apj, 837, 181, 1610.04862

\bibitem[{{Falck} {et~al.}(2012){Falck}, {Neyrinck}, \&
  {Szalay}}]{FalckEtal2012}
{Falck}, B.~L., {Neyrinck}, M.~C., \& {Szalay}, A.~S. 2012, \apj, 754, 126,
  1201.2353

\bibitem[{{Fall} \& {Romanowsky}(2018)}]{FallRomanowsky2018}
{Fall}, S.~M., \& {Romanowsky}, A.~J. 2018, \apj, 868, 133, 1808.02525

\bibitem[{{Gamow}(1952)}]{Gamow1952}
{Gamow}, G. 1952, Physical Review, 86, 251

\bibitem[{{Ganeshaiah Veena} {et~al.}(2019){Ganeshaiah Veena}, {Cautun},
  {Tempel}, {van de Weygaert}, \& {Frenk}}]{GaneshaiahEtal2019}
{Ganeshaiah Veena}, P., {Cautun}, M., {Tempel}, E., {van de Weygaert}, R., \&
  {Frenk}, C.~S. 2019, arXiv e-prints, 1903.06716

\bibitem[{{Ganeshaiah Veena} {et~al.}(2018){Ganeshaiah Veena}, {Cautun}, {van
  de Weygaert}, {Tempel}, {Jones}, {Rieder}, \& {Frenk}}]{GaneshaiahEtal2018}
{Ganeshaiah Veena}, P., {Cautun}, M., {van de Weygaert}, R., {Tempel}, E.,
  {Jones}, B.~J.~T., {Rieder}, S., \& {Frenk}, C.~S. 2018, \mnras, 481, 414,
  1805.00033

\bibitem[{{Gurbatov} {et~al.}(2012){Gurbatov}, {Saichev}, \&
  {Shandarin}}]{GurbatovEtal2012}
{Gurbatov}, S.~N., {Saichev}, A.~I., \& {Shandarin}, S.~F. 2012, Physics
  Uspekhi, 55, 223

\bibitem[{{Hahn} {et~al.}(2015){Hahn}, {Angulo}, \& {Abel}}]{HahnEtal2015}
{Hahn}, O., {Angulo}, R.~E., \& {Abel}, T. 2015, \mnras, 454, 3920, 1404.2280

\bibitem[{{Heavens} \& {Peacock}(1988)}]{HeavensPeacock1988}
{Heavens}, A., \& {Peacock}, J. 1988, \mnras, 232, 339

\bibitem[{Hidding(2019)}]{Hidding2019}
Hidding, J. 2019, PhD thesis, University of Groningen

\bibitem[{{Hidding} {et~al.}(2019){Hidding}, {van de Weygaert}, \&
  {Vegter}}]{HiddingEtal2019}
{Hidding}, J., {van de Weygaert}, R., \& {Vegter}, G. 2019, in prep

\bibitem[{Hidding {et~al.}(2012)Hidding, van~de Weygaert, Vegter, Jones, \&
  Teillaud}]{HiddingEtal2012}
Hidding, J., van~de Weygaert, R., Vegter, G., Jones, B.~J., \& Teillaud, M.
  2012, in Proceedings of the Twenty-eighth Annual Symposium on Computational
  Geometry, SoCG '12 (New York, NY, USA: ACM), 421--422

\bibitem[{Hoyle(1949)}]{Hoyle1949}
Hoyle, F. 1949, in {Problems of Cosmical Aerodynamics}, ed. J.~Burgers \&
  H.~van~de Hulst (Dayton, Ohio: Central Air Documents Office)

\bibitem[{{Jasche} \& {Lavaux}(2015)}]{JascheLavaux2015}
{Jasche}, J., \& {Lavaux}, G. 2015, \mnras, 447, 1204, 1402.1763

\bibitem[{{Joachimi} {et~al.}(2015){Joachimi}, {Cacciato}, {Kitching},
  {Leonard}, {Mandelbaum}, {Sch{\"a}fer}, {Sif{\'o}n}, {Hoekstra}, {Kiessling},
  {Kirk}, \& {Rassat}}]{JoachimiEtal2015}
{Joachimi}, B. {et~al.} 2015, \ssr, 193, 1, 1504.05456

\bibitem[{{Jones} \& {van de Weygaert}(2009)}]{JonesVdw2009}
{Jones}, B., \& {van de Weygaert}, R. 2009, Astrophysics and Space Science
  Proceedings, 8, 467, 0802.3599

\bibitem[{{Kitaura} \& {En{\ss}lin}(2008)}]{KitauraEnsslin2008}
{Kitaura}, F.~S., \& {En{\ss}lin}, T.~A. 2008, \mnras, 389, 497, 0705.0429

\bibitem[{{Kraljic} {et~al.}(2019){Kraljic}, {Pichon}, {Dubois}, {Codis},
  {Cadiou}, {Devriendt}, {Musso}, {Welker}, {Arnouts}, {Hwang}, {Laigle},
  {Peirani}, {Slyz}, {Treyer}, \& {Vibert}}]{KraljicEtal2019}
{Kraljic}, K. {et~al.} 2019, \mnras, 483, 3227, 1810.05211

\bibitem[{{Laigle} {et~al.}(2015){Laigle}, {Pichon}, {Codis}, {Dubois}, {Le
  Borgne}, {Pogosyan}, {Devriendt}, {Peirani}, {Prunet}, {Rouberol}, {Slyz}, \&
  {Sousbie}}]{LaigleEtal2013}
{Laigle}, C. {et~al.} 2015, \mnras, 446, 2744, 1310.3801

\bibitem[{{Liao} {et~al.}(2017){Liao}, {Gao}, {Frenk}, {Guo}, \&
  {Wang}}]{LiaoEtal2017}
{Liao}, S., {Gao}, L., {Frenk}, C.~S., {Guo}, Q., \& {Wang}, J. 2017, \mnras,
  470, 2262, 1610.07592

\bibitem[{{Libeskind} {et~al.}(2012){Libeskind}, {Hoffman}, {Knebe},
  {Steinmetz}, {Gottl{\"o}ber}, {Metuki}, \& {Yepes}}]{LibeskindEtal2012}
{Libeskind}, N.~I., {Hoffman}, Y., {Knebe}, A., {Steinmetz}, M.,
  {Gottl{\"o}ber}, S., {Metuki}, O., \& {Yepes}, G. 2012, \mnras, 421, L137,
  1201.3365

\bibitem[{{Ludlow} \& {Porciani}(2011)}]{LudlowPorciani2011}
{Ludlow}, A.~D., \& {Porciani}, C. 2011, \mnras, 413, 1961, 1011.2493

\bibitem[{{Marsh}(2016)}]{Marsh2016}
{Marsh}, D.~J.~E. 2016, \physrep, 643, 1, 1510.07633

\bibitem[{{Monaco} {et~al.}(2002){Monaco}, {Theuns}, \&
  {Taffoni}}]{MonacoEtal2002}
{Monaco}, P., {Theuns}, T., \& {Taffoni}, G. 2002, \mnras, 331, 587,
  arXiv:astro-ph/0109323

\bibitem[{{Neyrinck}(2016{\natexlab{a}})}]{Neyrinck2016iau}
{Neyrinck}, M.~C. 2016{\natexlab{a}}, in IAU Symposium, Vol. 308, The Zeldovich
  Universe: Genesis and Growth of the Cosmic Web, ed. R.~{van de Weygaert},
  S.~{Shandarin}, E.~{Saar}, \& J.~{Einasto}, 97--102, 1412.6114

\bibitem[{{Neyrinck}(2016{\natexlab{b}})}]{Neyrinck2016tet}
{Neyrinck}, M.~C. 2016{\natexlab{b}}, \mnras, 460, 816, 1510.03431

\bibitem[{{Neyrinck} {et~al.}(2018){Neyrinck}, {Hidding}, {Konstantatou}, \&
  {van de Weygaert}}]{NeyrinckEtal2018}
{Neyrinck}, M.~C., {Hidding}, J., {Konstantatou}, M., \& {van de Weygaert}, R.
  2018, Royal Society Open Science, 5, 171582, 1710.04509

\bibitem[{{Ozernoi}(1972)}]{Ozernoi1972}
{Ozernoi}, L.~M. 1972, \sovast, 15, 923

\bibitem[{{Peebles}(2018)}]{Peebles2018}
{Peebles}, P. 2018, in Astronomy in Focus, Vol.~1, XXXth IAU General Assembly,
  ed. T.~{Lago}, P.~{Tissera}, \& D.~{Obreschkow}, 15--19

\bibitem[{{Peebles}(1969)}]{Peebles1969}
{Peebles}, P.~J.~E. 1969, \apj, 155, 393

\bibitem[{{Pichon} \& {Bernardeau}(1999)}]{PichonBernardeau1999}
{Pichon}, C., \& {Bernardeau}, F. 1999, \aap, 343, 663, astro-ph/9902142

\bibitem[{{Platen} {et~al.}(2007){Platen}, {van de Weygaert}, \&
  {Jones}}]{PlatenEtal2007}
{Platen}, E., {van de Weygaert}, R., \& {Jones}, B.~J.~T. 2007, \mnras, 380,
  551, 0706.2788

\bibitem[{{Porciani} {et~al.}(2002){Porciani}, {Dekel}, \&
  {Hoffman}}]{PorcianiEtal2002}
{Porciani}, C., {Dekel}, A., \& {Hoffman}, Y. 2002, \mnras, 332, 325,
  astro-ph/0105123

\bibitem[{{Rindler-Daller} \& {Shapiro}(2012)}]{Rindler2012}
{Rindler-Daller}, T., \& {Shapiro}, P.~R. 2012, \mnras, 422, 135, 1106.1256

\bibitem[{{Schive} {et~al.}(2014){Schive}, {Chiueh}, \&
  {Broadhurst}}]{SchiveEtal2014}
{Schive}, H.-Y., {Chiueh}, T., \& {Broadhurst}, T. 2014, Nature Physics, 10,
  496, 1406.6586

\bibitem[{{Sciama}(1955)}]{Sciama1955}
{Sciama}, D.~W. 1955, \mnras, 115, 3

\bibitem[{{Springel}(2005)}]{Gadget2}
{Springel}, V. 2005, \mnras, 364, 1105, arXiv:astro-ph/0505010

\bibitem[{{Stewart} {et~al.}(2013){Stewart}, {Brooks}, {Bullock}, {Maller},
  {Diemand}, {Wadsley}, \& {Moustakas}}]{StewartEtal2013}
{Stewart}, K.~R., {Brooks}, A.~M., {Bullock}, J.~S., {Maller}, A.~H.,
  {Diemand}, J., {Wadsley}, J., \& {Moustakas}, L.~A. 2013, \apj, 769, 74,
  1301.3143

\bibitem[{{Sugerman} {et~al.}(2000){Sugerman}, {Summers}, \&
  {Kamionkowski}}]{SugermanEtal2000}
{Sugerman}, B., {Summers}, F.~J., \& {Kamionkowski}, M. 2000, \mnras, 311, 762,
  astro-ph/9909266

\bibitem[{{Trowland} {et~al.}(2013){Trowland}, {Lewis}, \&
  {Bland-Hawthorn}}]{TrowlandEtal2013}
{Trowland}, H.~E., {Lewis}, G.~F., \& {Bland-Hawthorn}, J. 2013, \apj, 762, 72,
  1201.6108

\bibitem[{{Vitvitska} {et~al.}(2002){Vitvitska}, {Klypin}, {Kravtsov},
  {Wechsler}, {Primack}, \& {Bullock}}]{VitvitskaEtal2002}
{Vitvitska}, M., {Klypin}, A.~A., {Kravtsov}, A.~V., {Wechsler}, R.~H.,
  {Primack}, J.~R., \& {Bullock}, J.~S. 2002, \apj, 581, 799, astro-ph/0105349

\bibitem[{{von Weizs{\"a}cker}(1951)}]{Wiseguy1951}
{von Weizs{\"a}cker}, C.~F. 1951, \apj, 114, 165

\bibitem[{{Wang} \& {White}(2007)}]{WangWhite2007}
{Wang}, J., \& {White}, S.~D.~M. 2007, \mnras, 380, 93, astro-ph/0702575

\bibitem[{{Wang} {et~al.}(2014){Wang}, {Szalay}, {Arag{\'o}n-Calvo},
  {Neyrinck}, \& {Eyink}}]{WangEtal2014}
{Wang}, X., {Szalay}, A., {Arag{\'o}n-Calvo}, M.~A., {Neyrinck}, M.~C., \&
  {Eyink}, G.~L. 2014, \apj, 793, 58, 1309.5305

\bibitem[{{White}(1984)}]{White1984}
{White}, S.~D.~M. 1984, \apj, 286, 38

\bibitem[{{Zel'dovich}(1970)}]{Zeldovich1970}
{Zel'dovich}, Y.~B. 1970, \aap, 5, 84

\end{thebibliography}

\end{document}